\documentclass[11pt,a4paper]{article}
\pdfoutput=1
\usepackage{jcappub}

\usepackage{epsfig}
\usepackage{url}
\usepackage{hyperref}

\usepackage[normalem]{ulem}

\usepackage{latexsym}
\usepackage{epsfig}
\usepackage{amsmath}
\usepackage{amssymb}
\usepackage{graphicx}
\usepackage{verbatim}
\usepackage{subfig}

\newcommand{\lb}{\left[}
\newcommand{\rb}{\right]}

\newcommand{\ba}{\begin{eqnarray}}
\newcommand{\ea}{\end{eqnarray}}
\newcommand{\be}{\begin{equation}}
\newcommand{\ee}{\end{equation}}
\newcommand{\bk}{{\bf k}}

\newcommand{\bn}{{\bf n}}

\newcommand{\la}{\lambda}

\newcommand{\Lag}{\mathcal{L}}

\newcommand{\HH}{{\cal H}}
\newcommand{\bv}{{\bf v}}

\usepackage[usenames,dvipsnames]{xcolor}

\newcommand{\du}[2]{_{#1}^{\phantom{#1} #2}}

\usepackage[all]{xy} 

\newcommand{\hiclass}{{\tt hi\_class}}

\title{Gravity at the horizon: on relativistic effects, CMB-LSS correlations and ultra-large scales in Horndeski's theory}

\author[a,b,c]{Janina Renk}
\author[b,a]{Miguel Zumalac\'arregui}
\author[d]{Francesco Montanari}
\affiliation[a]{Institut f\"ur Theoretische Physik, Ruprecht-Karls-Universit\"at Heidelberg, \\ Philosophenweg 16, 69120 Heidelberg, Germany}
\affiliation[b]{Nordita, KTH Royal Institute of Technology and Stockholm University \\ Roslagstullsbacken 23, SE-106 91 Stockholm, Sweden}
\affiliation[c]{The Oskar Klein Centre for Cosmoparticle Physics, Stockholm University \\  AlbaNova University Center, SE-106 91 Stockholm, Sweden}
\affiliation[d]{Physics Department, University of Helsinki and Helsinki Institute of Physics, \\
P.O. Box 64, 00014, University of Helsinki, Finland}

\emailAdd{renk@thphys.uni-heidelberg.de}
\emailAdd{miguel.zumalacarregui@nordita.org}
\emailAdd{francesco.montanari@helsinki.fi}

\abstract{
We address the impact of consistent modifications of gravity on the largest observable scales, focusing on relativistic effects in galaxy number counts and the cross-correlation between the matter large scale structure (LSS) distribution and the cosmic microwave background (CMB).
Our analysis applies to a very broad class of general scalar-tensor theories encoded in the Horndeski Lagrangian and is fully consistent on linear scales, retaining the full dynamics of the scalar field and not assuming quasi-static evolution.
As particular examples we consider self-accelerating Covariant Galileons, Brans-Dicke theory and parameterizations based on the effective field theory of dark energy, using the \hiclass\, code to address the impact of these models on relativistic corrections to LSS observables.
We find that especially effects which involve integrals along the line of sight (lensing convergence, time delay and the integrated Sachs-Wolfe effect -- ISW) can be considerably modified, and even lead to $\mathcal{O}(1000\%)$ deviations from General Relativity in the case of the ISW effect for Galileon models, for which standard probes such as the growth function only vary by $\mathcal{O}(10\%)$.
These effects become dominant when correlating galaxy number counts at different redshifts and can lead to $\sim 50\%$ deviations in the total signal that might be observable by future LSS surveys.
Because of their integrated nature, these deep-redshift cross-correlations are sensitive to modifications of gravity even when probing eras much before dark energy domination.
We further isolate the ISW effect using the cross-correlation between LSS and CMB temperature anisotropies and use current data to further constrain Horndeski models.
Forthcoming large-volume galaxy surveys using multiple-tracers will search for all these effects, opening a new window to probe gravity and cosmic acceleration at the largest scales available in our universe.
}

\date{\today}

\begin{document}

\maketitle

\section{Introduction}

The mechanism behind cosmic acceleration is one of the deepest mysteries confronted by modern science and might hold key information for our understanding of fundamental physics. One possibility for such a mechanism is a breakdown of Einstein's General Relativity (GR) on cosmological scales due to the presence of additional degrees of freedom \cite{Clifton:2011jh}.
Independently of possible connections with the phenomenon of cosmic acceleration, current and forthcoming cosmological observations will allow us to test gravity through many effects predicted by alternative theories using observations of the universe's expansion and the formation of large scale structure (LSS) \cite{Weinberg:2012es,Ade:2015rim,Koyama:2015vza}.
The area and depth covered by forthcoming surveys \cite{Laureijs:2011mu,Maartens:2015mra,Abell:2009aa} will extend this search for anomalous gravitational effects on increasingly larger volumes, eventually reaching into ultra-large scales comparable to the Hubble radius.

The Newtonian description of gravity becomes inaccurate on large cosmological scales, leaving room for relativistic effects that are otherwise negligible \cite{Yoo:2009au,Bonvin:2011bg,Challinor:2011bk}.\footnote{We will follow the convention and refer to relativistic effects as those which are sub-dominant on scales much smaller than the Hubble horizon (see \autoref{tab:rel_eff} and \autoref{sec:rel_eff}). However, we note that many effects that are often referred to as ``Newtonian'', such as redshift-space distortions and gravitational lensing, are also fully relativistic in nature. Similarly, we will adopt the convention of referring to Einstein's theory of gravity as General Relativity, despite the fact that all the theories we are considering obey the same principles and only distinguish themselves through the presence of additional degrees of freedom.}
Gravitational information takes cosmological timescales to travel across such vast distances, making the system's relaxation time comparable to its characteristic evolution rate.
Distortions of the light-cone caused by gravitational potentials can also affect the observables.
This and other effects have to be taken into account, at least in principle, and a growing body of literature is addressing the impact of these effects in LSS observations, as well as devising strategies to detect them with the next generation of galaxy surveys \cite{Bonvin:2013ogt,Irsic:2015nla,Montanari:2015rga,Alonso:2015sfa,Fonseca:2015laa,Cardona:2016qxn}.

Relativistic corrections to cosmological observables can provide new means to test gravity.
Ultra-large scales are interesting not only because these corrections become more important, but offer a number of advantages for cosmological tests of gravity.
The largest scales observable today have crossed the Hubble horizon at the onset of cosmic acceleration and are therefore attuned to the energy scale of any additional degree of freedom related to this phenomenon.
Such large scales are very well described by the linear theory and are the best bet to avoid screening mechanisms that strongly suppress the modifications of gravity due to non-linear effects \cite{Joyce:2014kja}. These screening mechanisms are predicted in many theories of gravity and can be very efficient in erasing the observational signatures on small and intermediate scales.
Finally, relativistic effects have a direct dependence on the metric perturbations and their evolution, which depends critically on the theory of gravity.

One of the fundamental observables in galaxy surveys are the Galaxy Number Counts (GNC's), which can be studied statistically through $n$-point correlation functions. The theoretical predictions can incorporate all relativistic effects. The GNC's depend only on directly observable quantities (angles and redshifts of galaxies), therefore angle and redshift dependent spectra can be consistently related to observations on arbitrary angular scales and redshift slicings without assuming a fiducial background expansion, thus offering considerable advantages over correlation functions in comoving space. In this sense, GNC angular statistics are the analogue of cosmic microwave background (CMB) spectra, with additional redshift dependence to take into account the 3-dimensional nature of the galaxy distribution.
Future surveys such as DESI, Euclid, SKA and the LSST \cite{Levi:2013gra,Laureijs:2011mu,Maartens:2015mra,Abell:2009aa} are expected to provide competitive measurements of such observables \cite{DiDio:2013sea}.

Combining different observables is also a promising way to constrain gravity. Galaxy and cosmic microwave background cross-correlations probe the late-time evolution of Bardeen potentials through the CMB integrated Sachs-Wolfe (ISW) effect \cite{Boughn:2003yz,Ade:2013dsi,Schmidt:2007vj,Ho:2008bz,Giannantonio:2012aa,Ferraro:2014msa}.
While non-linearities affect correlation functions on small scales, large enough angular scales can be studied using linear theory and provide valuable information about the evolution of Dark Energy.
Correlation of CMB data with future large scale surveys are expected to bring substantial improvements in discerning different Dark Energy models \cite{Majerotto:2015bra}.
CMB-LSS data have already been used to set constraints on alternative theories of gravity such as $f(R)$  \cite{PhysRevD.75.044004, PhysRevD.73.123504} and massive bigravity \cite{PhysRevD.91.084046}. It has also been discussed that Covariant Galileons predict an anti-correlation between galaxies and the temperature anisotropies in the CMB \cite{Barreira:2014jha}, which is in contradiction to the prediction of $\Lambda$-GR and the latest analysis by Ferraro et al. \cite{Ferraro:2014msa} using data from the WISE survey \cite{Wright:2010qw}.

Addressing the consequences of modified gravity on scales comparable to the Hubble radius requires going beyond the usual approximations valid on small scales. Most studies of LSS formation in alternative theories have been performed under the assumption of quasi-static (QS) evolution, in which the dynamics of the field are neglected \cite{DeFelice:2011hq,Amendola:2012ky}. This approximation has been shown to be consistent on small enough linear scales on which the dynamics have relaxed to the equilibrium values, with the regime of validity characterized by the sound horizon of the new degrees of freedom \cite{Sawicki:2015zya}. For this reason, the QS approximation is generally not consistent on ultra-large scales and it is necessary to solve the full system, either from a concrete theory or from an effective formulation such as the effective field theory of dark energy (EFT-DE) \cite{Gubitosi:2012hu,Bloomfield:2012ff,Gleyzes:2014rba}.
A particularly transparent application of the EFT-DE has been developed in the context of Horndeski's theory, the most general local, Lorentz-invariant, scalar-tensor theory described by second order equations of motion. This formalism, due to Bellini and Sawicki, employs properties of gravity instead of geometric quantities to parameterize cosmological perturbations in scalar-tensor theories \cite{Bellini:2014fua} and has been implemented in the \hiclass\, code \cite{Zumalacarregui:2016pph}.\footnote{\url{www.hiclass-code.net}}

In our analysis we investigate the modifications to the galaxy number counts in a very broad class of theories encompassed in the Horndeski Lagrangian.
Our analysis comprehends both, the best fit Galileon models of Barreira et al. \cite{Barreira:2014jha} as an ab initio model, as well as the parameterization of Horndeski proposed by Bellini \& Sawicki \cite{Bellini:2014fua} to separately control the background and the different properties that affect the perturbations.
We study systematically the dependence of each term contributing to the GNC's on the gravity parameters entering our models.
We also split all the contributions according to their scale dependence and physical interpretations.
The GNC's at redshifts from before Dark Energy domination can also have imprints from modified gravity if the signal is dominated by effects which are integrated along the line of sight. These effects like, e.g., lensing convergence, are affected by modifications of gravity on their way to the observer. We investigated this feature for parameterized Horndeski models.
Finally, we discuss some of the most remarkable signatures of modified gravity on the galaxy-CMB temperature anisotropy correlation on large angular scales.
We use the results for the measurements of the amplitude of the galaxy-CMB correlation from Ferraro et al. \cite{Ferraro:2014msa} to constrain the Bellini-Sawicki parameterization of Horndeski.
Our work expands on previous studies of modified gravity on ultra-large scales based on parameterizations of the solutions not relying on a particular theory \cite{Lombriser:2013aj,Baker:2015bva}.
This analysis is meant to serve as a general guide for future tests of gravity through galaxy surveys on very large scales.

In \autoref{sec:rel_eff} we review the derivation of the full relativistic galaxy number counts and discuss the scale dependence and detection prospects of the several effects. In \autoref{sec:ScalarGrav} we provide a short recap of scalar-tensor theories of gravity, focusing on the models we are using in our analysis: Covariant Galileons and the parameterized Horndeski's theory of gravity. In \autoref{sec:GNC} we investigate how the change of the matter and metric perturbations of these models influence the GNC's and the single contributions to it.
In \autoref{sec:BinCrossCorr} we address the signatures of the paramterized Horndeski's theory on the GNC's for correlations of high redshift bins from eras before Dark Energy domination.
Finally, in \autoref{sec:CMB-LSS}, we use the cross-correlation between CMB temperature anisotropies and large scale structure to constrain the parameter space of the parameterized Horndeski theory of gravity.
We conclude with summarizing and discussing our results in \autoref{sec:Concl}.

For the busy reader who is familiar with relativistic effects in GNC's we recommend to skip \autoref{sec:rel_eff}. Similarly, readers familiar with Horndeski's theory and the effective field theory of Dark Energy can safely skip \autoref{sec:ScalarGrav}.
Readers whose is main interest is on methods to test gravity using different effects can jump to \autoref{sec:BinCrossCorr} and \autoref{sec:CMB-LSS}.

\section{Relativistic effects in galaxy number counts}
\label{sec:rel_eff}
Let us denote by $N(\bn,z)$ the total number of galaxies within a given redshift and angular bin, where $-\bn$ is the direction of observation ($\bn$ being the photon's propagation direction) and $z$ is the redshift, and by $V(\bn,z)$ the volume integrated within the redshift and angular bin.
Then $n_{\rm g}(\bn,z)=d N(\bn,z)/dz/d\Omega$ and $\nu(\bn,z)=d V(\bn,z)/dz/d\Omega$ are the number density of galaxies and the volume density, respectively, per redshift $z$ and solid angle $\Omega$.
Galaxy number counts are defined as the contrast with respect to the angle averaged value $\langle n_{\rm g} \rangle (z)$:
\begin{equation} \label{eq:GNC}
\Delta(\bn,z) = \frac{n_{\rm g}(\bn,z)-\langle n_{\rm g} \rangle (z)}{\langle n_{\rm g} \rangle (z)}.
\end{equation}
This observable is used to compute correlation functions to be compared to theoretical models.
In the context of perturbation theory~\cite{Ma:1995ey,Durrer:2007zz}, density fluctuations (and similarly for other perturbations) are usually described as a subtraction between the true galaxy density, $n_{\rm g}(\bn,z)$, defined on the true space-time and its value $\bar n_{\rm g}(\bar z)$ on a homogeneous and isotropic Friedmann-Lema\^itre-Robertson-Walker (FLRW) background:
\begin{equation} \label{eq:delta_g}
\delta_{\rm g}(\bn,z) = \frac{n_{\rm g}(\bn,z)-\bar n_{\rm g} (\bar z)}{\bar n_{\rm g} (\bar z)} \;.
\end{equation}
Note that even though perturbations are defined on the background space-time for simplicity of notation we indicate the dependence on the observed coordinates, as differences are second-order small.
Furthermore, spatial averages of perturbations vanish, such that $\bar n_{\rm g} = \langle n_{\rm g} \rangle$.
Let us introduce the galaxy density per comoving volume with $\rho_{\rm g}(\bn,z)=n_{\rm g}(\bn,z)/\nu(\bn,z)$.
In the following we consider linear perturbation theory which allows to write the volume density in terms of its background value plus a perturbation, $\nu(\bn,z) = \bar\nu(z)+\delta\nu(\bn,z)$.
Following \cite{Bonvin:2011bg}, we write eq.~(\ref{eq:GNC}) to linear order as
\begin{equation} \label{eq:GNC_p}
\Delta(\bn,z) = \frac{\rho_{\rm g} \left( \bn , z \right) - \bar \rho_{\rm g} \left( z \right) }{\bar \rho_{\rm g} \left( z \right)} + \frac{\delta \nu \left( \bn , z \right)}{\bar \nu \left( z \right)} \;.
\end{equation}
It can be shown that the two terms on the right hand side are separately gauge invariant.
Instead, the $\delta_{\rm g}$ perturbation, defined in eq.~(\ref{eq:delta_g}) through the fictitious background coordinates, depends on the specific gauge choice.
The first term of eq.~(\ref{eq:GNC_p}) is the subtraction between the full density and its background value both evaluated at the observed redshift.
To compute it with perturbation theory, we want to replace this difference with a subtraction between two space-times, where the background density is then evaluated at the background redshift.
To do this, we expand the redshift as $z=\bar z+\delta z$.
The first term on the right hand side of eq.~(\ref{eq:GNC_p}) can be written in terms of the redshift perturbations $\delta z$ by Taylor expanding around $\bar z$:
\begin{equation}
  \frac{\rho_{\rm g}(\bn,z)-\bar\rho_{\rm g}(z)}{\bar\rho_{\rm g}(z)}
  = \delta_{\rm g}(\bn,z) - \frac{d\bar\rho_{\rm g}(\bar z)}{d\bar z} \frac{\delta z(\bn,z)}{\bar\rho_{\rm g}(\bar z)} \;.
\end{equation}
Similarly, we can also expand the direction of observation, $\bn = \bar\bn + \delta\bn$, appearing in the volume perturbation.

Hence, given the FLRW background evolution eq.~(\ref{eq:GNC_p}) is determined in terms of the density perturbation, $\delta_{\rm g}$ (well-known from standard perturbation theory, see \cite{Ma:1995ey, Durrer:2007zz}), the redshift perturbation, $\delta z$, and the volume perturbation, $\delta\nu$.
These can be computed by solving the geodesic equations as showed, e.g., in \cite{Yoo:2009au,Yoo:2010ni,Bonvin:2011bg,Challinor:2011bk,Jeong:2011as,Jeong:2012nu}.
As the observable $\Delta(\bn,z)$ is gauge-invariant, the computation can be carried out in any gauge.
We consider the longitudinal Newtonian gauge:
\begin{equation} \label{eq:metric}
ds^2 = a^2 \left[ - \left( 1 + 2 \Psi \right) d\tau^2 + \left( 1 - 2 \Phi\right) \gamma_{ij} dx^i dx^j \right] \;,
\end{equation}
where $a$ is the scale factor, $\Psi$ and $\Phi$ are the Bardeen potentials, and the spatial part of the metric is
\begin{equation}
 \gamma_{ij} dx^i dx^j = \left[ dr^2 + r^2 \left( d\theta^2 + \sin^2 \theta d\varphi^2\right) \right] \;.
\end{equation}
Here and throughout the paper we use natural units in which $c=1$.

Galaxy number counts $\Delta_{\rm obs} ({\bf n}, z)$ are given by the sum of the following terms
\begin{eqnarray}
\Delta_{\delta}({\bf n}, z)  &=& b(z)\;\delta_{\rm co}\left(r(z)\bn,\tau(z)\right) \,,
\label{eq:D_delta}\\
\Delta_{\rm rsd}({\bf n}, z)  &=& \frac{1}{\HH(z)}\partial_r(\bv\cdot\bn) \,,
\label{eq:D_rsd} \\
\Delta_{\rm \kappa}({\bf n}, z)  &=& -\frac{(2-5s(z))}{2}\int_0^{r(z)}dr\frac{r(z)-r}{r(z)r}\Delta_2(\Phi+\Psi) \,,
\label{eq:D_kappa} \\
\Delta_{\rm dop}({\bf n}, z)  &=& \left[\frac{\HH'}{\HH^2}+\frac{2-5s(z)}{r\HH} + 5s(z)- f_{\rm evo}(z)\right](\bv\cdot\bn)+
\nonumber \\ &&
\quad + \left[3-f_{\rm evo}(z)\right]\HH\Delta^{-1} (\nabla\cdot\bv) \,,
\label{eq:D_v} \\[2pt]
\Delta_{\rm lp}({\bf n}, z)  &=&  (5s(z)-2)\Phi+\Psi +\HH^{-1}\Phi'
\nonumber \\ &&
+ \left[\frac{\HH'}{\HH^2}+\frac{2-5s}{r\HH}+ 5s(z)- f_{\rm evo}(z)\right]\Psi \,,
 \label{eq:D_pot_loc} \\
\Delta_{\rm td}({\bf n}, z)  &=&  \frac{2-5s}{r(z)}\int_0^{r(z)}dr(\Phi+\Psi) \,,
 \label{eq:D_pot_std} \\
\Delta_{\rm ISW}({\bf n}, z)  &=&   \label{eq:D_pot_isw}
 \left[\frac{\HH'}{\HH^2}+\frac{2-5s}{r\HH}+ 5s(z)- f_{\rm evo}(z)\right]\int_0^{r(z)}dr(\Phi'+\Psi') \,.
\end{eqnarray}
Here $\bv$ is the velocity perturbation in longitudinal gauge, $\HH = aH$ is the conformal Hubble parameter and $\Delta_2$ is the Laplacian on the sphere.
The  term $\Delta^{-1}(\nabla\cdot\bv)$ is the velocity potential (in Fourier space it reads $-V(k)/k$, given the $k$-mode $V(k)$ of the velocity field) coming from the gauge transformation relating the density perturbation in the Newtonian gauge to the one in the synchronous gauge comoving with Dark Matter $\delta_{\rm co}$. This is the one entering in the Poisson equation \cite{Wands:2009ex,Bertacca:2015mca} and that should then be multiplied by the (linear) bias factor $b(z)$.
All quantities not integrated are evaluated at conformal time $\tau(z)$ and at position $r(z)(-\bn)=(\tau_0-\tau(z))(-\bn)$. A prime indicates a derivative w.r.t. conformal time.
We also introduce the magnification bias $s(z)$ and evolution bias $f_{\rm evo}(z)$ terms.
In particular, the latter takes into account deviations from the conservation of the background comoving galaxy number density $a^3 \bar \rho_{\rm g}=$const., assumed in the previous equations.

\begin{table}[t]
\makebox[\linewidth]{
 \begin{tabular}{llcccc  c}
 & Effect & scaling & $z$-dep. & bias dep.  & Eq. & Notes \\ \hline \hline
$\Delta_\delta$  & Intrinsic clustering & $\left(\frac{k}{\HH}\right)^2\delta g_{\mu\nu}$ & Local & $b$ & (\ref{eq:D_delta}) & $z_i\approx z_j$
\\
$\Delta_{\rm rsd}$  & Kaiser RSD & $\left(\frac{k}{\HH}\right)^2\delta g_{\mu\nu}$ & Local & - & (\ref{eq:D_rsd}) & small $\Delta z$
\\ \hline
$\Delta_{\kappa}$  & Lensing convergence & $ \left(\frac{k}{\HH}\right)^2\delta g_{\mu\nu}$ & Integrated & $s$ & (\ref{eq:D_kappa}) & $z_i\neq z_j$ \\ \hline
$\Delta_{\rm d}$  & Doppler effects & $\left(\frac{k}{\HH}\right)\delta g_{\mu\nu}$ & Local & $s,f_{\rm evo}$ & (\ref{eq:D_v}) & multitracers
\\ \hline
$\Delta_{\rm lp}$  & Local potentials  & $\delta g_{\mu\nu}$ & Local & $s,f_{\rm evo}$ & (\ref{eq:D_pot_loc}) &  multitracers \\
$\Delta_{\rm td}$  & Shapiro time delay  & $\delta g_{\mu\nu}$ & Integrated & $s$ & (\ref{eq:D_pot_std}) & multitracers \\
$\Delta_{\rm ISW}$  & Integrated Sachs-Wolfe  & $\delta g_{\mu\nu}$ & Integrated & $s,f_{\rm evo}$ & (\ref{eq:D_pot_isw}) & multitracers
\\
  \hline\hline
 \end{tabular}}
\caption{Summary of relativistic effects including their dependence on scale, redshift and bias. For some effects it has been pointed out which configurations and observables are most sensitive. Relativistic effects have been shown to be detectable through multi-tracer observations with future surveys \cite{Alonso:2015sfa,Fonseca:2015laa}, but they have not been addressed independently. Note that the ISW effect can be isolated by considering CMB-LSS cross-correlations, cf. \autoref{sec:CMB-LSS}.
}
  \label{tab:rel_eff}
\end{table}

The first three contributions $\Delta_{\delta}$ (\emph{intrinsic clustering}), $\Delta_{\rm rsd}$ (Kaiser \emph{redshift-space distortions}) and $\Delta_{\rm \kappa}({\bf n}, z)$ (\emph{lensing convergence})\footnote{Note that the fully relativistic expression for convergence $\kappa$ should be obtained by solving the Sachs equation in a perturbed FLRW universe \cite{Bonvin:2014owa}, which then involve other terms than gravitational lensing only, but we include them into the definition of the other effects. It is also worth recalling that $\kappa$ is a gauge-dependent quantity \cite{Yoo:2014kpa} and, as such, not observable by its own in a relativistic context.} are proportional to two spatial derivatives of the Bardeen potentials.
The velocity terms $\Delta_{\rm d}$ (which we call \emph{Doppler} effects) are instead proportional to one spatial derivative of Bardeen potentials, while the remaining terms depend directly on them. We have defined the \emph{integrated Sachs-Wolfe} effect (ISW) by those terms $\Delta_{\rm ISW}({\bf n}, z)$ proportional to $\int_0^{r(z)}dr(\Phi'+\Psi')$ and the \emph{Shapiro time-delay} by terms $\Delta_{td}({\bf n}, z)$ proportional to $\int_0^{r(z)}dr(\Phi+\Psi)$. We refer to local contribution depending on the Bardeen potentials and their time derivatives as \emph{local potential} effects $\Delta_{\rm lp}({\bf n}, z)$. These terms are summarized in \autoref{tab:rel_eff} and further described in \autoref{sec:transfer}.
In Fourier space, Doppler and potential terms are suppressed by a factor $\HH/k$ and $\left(\HH/k\right)^2$, respectively, compared to the leading terms.
Therefore, they are negligible on scales much smaller than the horizon $k\gg\HH$, while they are relevant on large scales $k\sim\HH$.
Furthermore, terms integrated along the line of sight (lensing convergence, ISW effect and Shapiro time-delay) are negligible on small and intermediate radial scales where galaxy clustering is dominated by local contributions.
However, they can be relevant if ultra-large radial scale correlations are taken into account in the analysis.

We summarize the scale dependence of the different effects in \autoref{tab:rel_eff}, which serves as a guide for detection studies.
Local terms such as RSD, Doppler and local potentials terms are enhanced when correlations at the same redshift are taken into account, while they decay on large redshift separations, where instead terms integrated along the line of sight (lensing, Shapiro time-delay and ISW effect) are more relevant.
Because of the factor's $(k/\HH)$ suppression previously discussed, only the intrinsic clustering, RSD and lensing effects are relevant for single tracer analysis, while the other terms (especially the dominant Doppler effects) may be detected combining different large-scale structure tracers \cite{Alonso:2015sfa,Fonseca:2015laa}.
Let us also note that a clear detection of these effects, however, also relies on a good modelling of observational effects not treatable in perturbation theory such as magnification and evolution bias, depending on the source luminosity function.

The results discussed so far do not assume Einstein's equations and are independent of the gravitational theories as long as galaxies follow geodesics,
\begin{equation}
\label{eq:gal_geo}
\bn \cdot \bv' + \HH \bn \cdot \bv - \partial_{r}\Psi=0 \;.
\end{equation}
This assumption can be violated in certain scalar-tensor theories of gravity \cite{Hui:2009kc}. Departures from geodesic motion can be very strong in theories such as chameleons or symmetrons \cite{Khoury:2003aq,Hinterbichler:2010es}, in which unscreened objects fall in the geodesics of (\ref{eq:metric}) while the trajectories of screened objects are determined by a different effective metric involving the scalar field.\footnote{More specifically, unscreened objects follow the geodesics of the Jordan frame metric, \ref{eq:metric}, while screened objects follow geodesics of the Einstein frame metric (as they are not sensitive to the modifications of gravity). The Einstein and Jordan frame metric are related by a field dependent rescaling $g_{\mu\nu}^{(E)}=C(\phi)g_{\mu\nu}$, where the conformal factor $C(\phi)$ depends on the theory.}
However, departures from geodesic motion (\ref{eq:gal_geo}) are very suppressed for many scalar-tensor theories of interest. This includes Covariant Galileons and other theories featuring the Vainshtein screening mechanism \cite{Vainshtein:1972sx}. In these cases the modified gravity effects enter only through the different relations between the gravitational potentials and the matter distribution, which will be affected by the additional degree of freedom.

Galaxy number counts can be used to estimate two-point statistics.
As motivated in \cite{DiDio:2013sea} angular power spectra are particularly suited to study relativistic effects as they depend on the directly observable multipoles $\ell$'s (encoding the angular dependence) and redshifts.
Being the angular spectra computed into redshift shells, we indicate the correlation between the bins at reference redshifts $z_i$ and $z_j$ by $C_{\ell}(z_i,z_j)$.
Furthermore, the transfer functions $\Delta_{\ell}^{W_i}$ depend in general on a window function $W(z_i)$ encoding, e.g., information about the galaxy distribution and redshift resolution.
The power spectrum of the primordial curvature perturbation $\mathcal{R}(\bk)$ is defined as
  \begin{equation}
    k^3 \langle \mathcal{R}(\bk)\mathcal{R}(\bk') \rangle = \delta_D(\bk-\bk') P_{\mathcal{R}}(k) \;,
    \end{equation}
  where $\delta_D(\bk-\bk')$ is the Dirac delta function, and we also introduce the adimensional spectrum\footnote{Our conventions are consistent with the unitary normalization of Fourier transforms.} $\mathcal{P}_{\mathcal{R}}(k)=\frac{k^3}{2\pi^2}P_{\mathcal{R}}(k)$.
The redshift and angle dependent power spectra are given by:
\begin{equation}
\label{eq:Cls}
C_{\ell}(z_i,z_j) = 4\pi \int \frac{dk}{k} \Delta_{\ell}^{W_i}(k) \Delta_{\ell}^{W_j}(k) \mathcal{P}_{\mathcal{R}}(k) \;,
\end{equation}
where the transfer functions $\Delta_{\ell}^{W_i}$ are given in \autoref{sec:transfer}.

Single tracer analysis are mostly sensitive to the density, redshift-space distortions and lensing effects \cite{Montanari:2015rga}.
Detection of other relativistic effects with single-tracer analysis applied on Euclid-like spectroscopic surveys was ruled out in \cite{Yoo:2013zga,Yoo:2013tc} (see however \cite{Raccanelli:2016avd}), while a marginal detection is expected in the context of multiple-tracer analysis in the plane-parallel approximation \cite{Yoo:2012se}.
Promising detection results have been obtained with multiple-tracer
analysis using full-sky angular power spectra~\cite{Alonso:2015sfa,Fonseca:2015laa}.
These studies rely on the fact that large scale observables depending on a (deterministic) galaxy bias, such as the local primordial non-Gaussianities parameter, can be measured with a strong reduction of cosmic variance \cite{Seljak:2008xr}.
Also symmetries in the correlation function can be used to isolate relativistic effects \cite{Bonvin:2015kuc,Gaztanaga:2015jrs,Irsic:2015nla,Raccanelli:2013dza}, opening another new window where our theory of gravity can be constrained.
We leave a detailed multi-tracer detection study through relativistic effects of deviations from GR based on Horndeski models as a future project.

\section{Scalar-tensor gravity}
\label{sec:ScalarGrav}

Alternative theories of gravity require the introduction of additional degrees of freedom or other departures from the minimal assumptions that lead to Einstein's theory. A simple scenario is to postulate an additional scalar field that interacts non-minimally with the metric.
In this context Horndeski's theory \cite{Horndeski:1974wa} provides a general framework encompassing scalar-tensor theories that are local, Lorentz-invariant and have their dynamics described by second order equations of motion.%
\footnote{This is a sufficient condition to avoid the introduction of additional, ghost degrees of freedom \cite{Woodard:2015zca}, although not a necessary one \cite{Zumalacarregui:2013pma,Gleyzes:2014dya}. Note that theories beyond Horndeski are actually \emph{more restricted} than Horndeski because of the less efficient Vainshtein mechanism \cite{Kobayashi:2014ida}, the lower effective field theory cutoff \cite{Pirtskhalava:2015nla} and the presence of ghosts when combined with Horndeski terms \cite{Langlois:2015cwa, Crisostomi:2016tcp}.}

It is given by the following action functional
\begin{equation}\label{eq:hornyLagrangian}
 S[g_{\mu\nu},\phi] = \int \sqrt{-g}\sum_{i=2}^{5} \Lag_i\,,
\end{equation}
where the different pieces can be written as:
\begin{eqnarray}
\Lag_2 &=&   G_2(X,\phi)\,, \label{LH2}
 \\[5pt]
\Lag_3 &=& -G_3(X,\phi) \Box \phi\,, \label{LH3}
 \\[5pt]
\Lag_4 &=& G_4(X,\phi) R + G_{4,X}\lb (\Box\phi)^2 - \phi_{;\mu\nu}\phi^{;\mu\nu} \rb\,, \label{LH4}
\\[5pt]
\Lag_5 &=& G_5(X,\phi) G_{\mu\nu}\phi^{;\mu\nu} - \frac{1}{6}G_{5,X}\big[(\Box\phi)^3 - 3(\Box\phi)\phi_{;\mu\nu}\phi^{;\mu\nu}
+ 2\phi\du{;\mu}{;\nu} \phi\du{;\nu}{;\la} \phi\du{;\la}{;\mu}\big]\,. \label{LH5}
\end{eqnarray}
Horndeski's theory has attracted considerable attention as a general framework that encompasses many theories of interest, including Brans-Dicke \cite{Brans:1961sx}, quintessence, kinetic gravity braiding \cite{Deffayet:2010qz}, Covariant Galileons \cite{Deffayet:2009wt,Deffayet:2009mn}, Gauss-Bonnet couplings \cite{Maselli:2015yva,Ezquiaga:2016nqo}, disformal gravity \cite{Zumalacarregui:2012us,Bettoni:2013diz} and some degravitation models \cite{Charmousis:2011bf,Martin-Moruno:2015bda}.

In order to tame the vast functional freedom of Horndeski's theory, several approaches have been devised to describe the dynamics of linear perturbations over a cosmological background \cite{Gubitosi:2012hu,Bloomfield:2012ff,Gleyzes:2013ooa}. A particularly clear formulation to capture linear structure formation in these models has been developed by Bellini and Sawicki \cite{Bellini:2014fua}.%
\footnote{For extensions to the non-linear regime see \cite{Bellini:2015wfa,Bellini:2015oua}.}
In their formalism, the perturbed dynamics are determined by the background contribution $\Omega_{de}$ and a set of four dimensionless functions that depend on time and represent distinct physical properties of the theory:
\begin{itemize}
 \item The \emph{kineticity}, $\alpha_K$, characterizes the standard kinetic term for the scalar degree of freedom and modulates the speed of sound for scalar perturbations. It receives contributions from all of the Horndeski functions (except a potential term $G_2(\phi)$) and is the only non-zero $\alpha$-parameter in models with only $G_2$ such as quintessence and K-essence. Note that in the limit of quasi-static evolution the dynamics become independent of the kineticity.

 \item The \emph{braiding}, $\alpha_B$, characterizes the non-diagonal part of the kinetic term, i.e. it causes second time derivatives of the gravitational potential to enter the equation of the scalar field and vice-versa. This \emph{kinetic mixing} between the metric and scalar degrees of freedom is a characteristic property of scalar-tensor theories \cite{Bettoni:2015wta}. $\alpha_B$ is sourced by $G_{3,X}$, $G_4$ and $G_5$ and is responsible for the scale dependence of the effective gravitational constant.

 \item The \emph{running}, $\alpha_M = \frac{d\log(M_*^2)}{d\log(a)}$, or time variation of the effective Planck mass $M_*$ generalizes the notion of an evolving gravitational constant. A non-zero value of $\alpha_M$ is equivalent to a violation of energy conservation (e.g. in the Einstein frame representation of Brans-Dicke theories). It is sourced by $G_{4}$ and $G_5$ and for generalized Brans-Dicke theories in which $G_4=f(\phi)$ satisfies $\alpha_M = -\alpha_B$.

 \item The \emph{tensor speed excess}, $\alpha_T$, captures the difference between the speed of light and the propagation speed of tensor modes on a cosmological background, $c_{GW} = 1 + \alpha_T$. This phenomenon implies a different causal structure for matter and gravity and induces an offset between the two scalar gravitational potentials. It occurs in theories with $G_{4,X}$ or $G_{5}$.

\end{itemize}
Fixing a background evolution for $H(t)$, the matter density parameter today, $\Omega_{\rm m}$, and the four functions $\alpha_i$ fully determines the evolution of large-scale structure providing a minimal set of functions able to describe all models consistent with Horndeski gravity at linear order in perturbation theory.
The first two functions ($\alpha_K$, $\alpha_B$) characterize the kinetic term for the scalar degree of freedom. Their interplay determines the \emph{braiding scale}
\begin{equation}\label{eq:braiding_scale}
\frac{k_{\text{B}}^{2}}{a^{2}H^{2}}\equiv \frac{\alpha_K + \frac{3}{2}\alpha_B^2}{\alpha_{\textrm{B}}^{2}}\left[\left(1-\frac{\Omega_{\text{m}}}{M_*^2}\right)(1+w_{X})+2\left(\alpha_{\textrm{M}}-\alpha_{\textrm{T}}\right)\right]+\frac{9}{2}\frac{\Omega_{\text{m}}}{M_*^2}\,,
\end{equation}
responsible for the scale-dependent growth predicted by scalar-tensor gravity in the quasi-static regime \cite{DeFelice:2011hq,Amendola:2012ky}.
The last two functions ($\alpha_M$, $\alpha_T$) can be thought of as genuine modifications of gravity, as they enter the evolution equation for tensor modes on a cosmological background and also introduce an offset between the two gravitational potentials \cite{Saltas:2014dha}.

In addition, we require stability of the perturbations on all scales. This amounts to demanding a positive scalar kinetic term $D\equiv \alpha_K+\frac{3}{2}\alpha_B>0$ and speed of sound
\begin{eqnarray} \label{eq:c_s}
c_{s}^{2}  = & \Bigg[ &
\left(2-\alpha_{\textrm{B}}\right)\left[6+\alpha_{\textrm{B}}+2\alpha_{\textrm{M}}-\alpha_{\textrm{T}}\left(2-\alpha_{\textrm{B}}\right)+3\left(w_{\text{de}}\Omega_{\text{de}}+w_{\textrm{m}}\Omega_{\textrm{m}}\right)\right]
\nonumber \\ &  &
-6\left(2+w_{\textrm{m}}\right)\frac{\Omega_{\textrm{m}}}{M_{*}^{2}}+\frac{2\dot{\alpha}_{\textrm{B}}}{H}
\Bigg]
({3\alpha_{\textrm{B}}^{2}+2\alpha_{\textrm{K}}})^{-1}>0
\,,
\end{eqnarray}
as well as similar conditions for tensor modes $M_*^2>0$ and $1+\alpha_T>0$. Note that the scalar speed of sound (\ref{eq:c_s}) also determines the sound horizon, below which the field can be considered to evolve quasi-statically \cite{Sawicki:2015zya}.

We will consider both,  Covariant Galileons, as an example for a self-consistent ab-initio model, as well as a parameterization of the $\alpha$-functions that determine the dynamics of cosmological perturbations and can be used to study their effects separately.

\subsection{Covariant Galileons}
\label{sec:CovGal}

The archetypal example of a modern Horndeski theory is the Covariant Galileon \cite{Deffayet:2009wt} as the extension of the flat-space Galileons \cite{Nicolis:2008in} to curved space-time and second-order equations. It is defined by
\begin{equation}
 G_2=c_1\phi - c_2 X\,,\; G_3 = \frac{c_3}{M^3}X\,,\;
 G_4 = \frac{M_p^2}{2} - \frac{c_4}{M^6}X^2\,,\; G_5 = \frac{3c_5}{M^9}X^2\,. \label{eq:cov_gal_action}
\end{equation}
We call \emph{cubic, quartic} and \emph{quintic Galileons} those models with Lagrangian given by $\mathcal{L}_2+\mathcal{L}_3$, $\mathcal{L}_2+\mathcal{L}_3+\mathcal{L}_4$ and $\mathcal{L}_2+\mathcal{L}_3+\mathcal{L}_4+\mathcal{L}_5$, respectively, when assuming (\ref{eq:cov_gal_action}).
In other words, this corresponds to setting both $c_4=c_5=0$, only $c_5=0$ or leaving all the $c_i$'s different from zero, respectively.
We use the normalization of the scalar field to set $c_2 = 1$ and will not consider the linear potential $c_1=0$ (see Ref. \cite{Barreira:2014jha} for further details).

The observational viability of this theory has been discussed in \cite{Barreira:2014jha} (see also \cite{Barreira:2012kk,Barreira:2013jma}), which also provides an excellent introduction into the dynamics of the model. The Covariant Galileon is able to fit the background expansion of the universe and the CMB spectra. However, these models predict a fairly large growth of matter perturbations due to both, the enhanced gravitational force and the larger value of $H_0$ needed to fit the background expansion \cite{Appleby:2012ba}. Including neutrino masses alleviates these effects and gives a better fit to data \cite{Barreira:2014jha}.

Note that Galileons are equipped with the Vainshtein screening mechanism, which hides the modifications of gravity on small scales due to the non-linear derivative self-interactions of the scalar degree of freedom.
Numerical simulations indicate that the screening becomes effective on sub-horizon scales for the Galileon models we will consider \cite{Barreira:2013eea,Li:2013tda}. However, this effect provides an additional motivation to consider ultra-large cosmological scales, in which the linear behaviour is an excellent approximation and no screening mechanisms are active.

\subsection{Brans-Dicke theory}

Another important example of a Horndeski model stemming from a covariant Lagrangian is given by Brans-Dicke theory. Because this is a very minimal theory and is not able to self-accelerate the universe, we have included only a brief discussion in \autoref{sec:BD_theory}.

\subsection{Parameterized Horndeski gravity}\label{sec:param_bell_sawi}

It is also possible to consider modifications of gravity not based on any specific Lagrangian by specifying a parameterization of the background expansion and the $\alpha$-functions. This approach is complementary to the study of specific models, as it allows one to control the different properties of gravity and the cosmological expansion independently.
We will adopt the parameterization proposed by Bellini and Sawicki \cite{Bellini:2014fua}
\begin{equation}\label{eq:propto_omega}
  \alpha_i(a) = c_i \Omega_{\rm de}(a) = c_i\frac{\rho_{\rm de}}{\rho_{\rm crit}} \,.
\end{equation}
The initial effective Planck mass $M_*$ is set to one in units of $M_p$. The $\alpha_i$'s and $c_i$'s correspond to kineticity ($\alpha_K$ and $c_K$), braiding ($\alpha_B$ and $c_B$), running ($\alpha_M$ and $c_M$) and tensor speed excess ($\alpha_T$ and $c_T$) respectively. General Relativity is recovered if all $\alpha_i$'s are equal to zero.
The proportionality of the $\alpha_i$'s to the energy density of Dark Energy ensures that we do not include early time modifications to the theory of gravity. Therefore high redshift phenomena, like the CMB, are not altered.
The parameterization chosen for the $\alpha$-functions coincides exactly with simple models such as Kinetic Gravity Braiding \cite{Deffayet:2010qz}.
In general it will not reproduce exactly all specific models (see section \ref{sec:gal_param} for a comparison with Galileons), but it has the advantage of providing in principle a consistent description of the Horndeski Lagrangian, extending other schemes usually adopted, such as parameterizations of the effective gravitational constant and the value of $\Psi-\Phi$ based on the quasi-static approximation (e.g. \cite{Baker:2015bva}).

To fully describe the evolution of large-scale structure, besides the $\alpha$-functions (characterizing the perturbations), we need to specify a background evolution. We will assume that the evolution of the scale factor is given by the standard Friedmann equation $H^2 = \frac{8\pi G}{3}\left(\rho_m + \rho_{\rm de}\right)$ sourced by matter and a Dark Energy component with
\begin{equation}\label{eq:expansion}
 \rho_{de}(a) = \rho_{\rm de,0} a^{3(1+w)}\,.
\end{equation}
This background expansion reproduces the results of standard gravity with a cosmological constant (if $w=-1$), therefore allowing us to focus on the novel effects at the level of linear perturbations.

The impact of this model in LSS observables has been addressed in \cite{Bellini:2015wfa}, including constraints on the $c_i$-functions using current data \cite{Bellini:2015xja}. In order to study the effects of each $\alpha$-function separately we run a series of Markov Chain Monte Carlo analyses varying one or two $c_i$'s only (see section \ref{sec:Param}).
Demanding stability of the perturbations on all scales (\ref{eq:c_s}) is very constraining on these models.%
\footnote{However, if $|c_s^2|\ll 1$ the instabilities will only occur on non-linear scales on which our formalism does not apply. This should hold true for models with the Vainshtein screening, for which standard gravity should be recovered on small scales. The Vainshtein mechanism has been argued to prevent instabilities \cite{Mortsell:2015exa, Aoki:2015xqa} that are ubiquitous in bigravity theories \cite{Konnig:2015lfa}.}

\subsubsection{Covariant Galileons in the parameterized language }
\label{sec:gal_param}
\begin{figure}
 \centering
  \begin{minipage}{0.49\linewidth}
    \centering
   {\includegraphics[width=\textwidth]
   {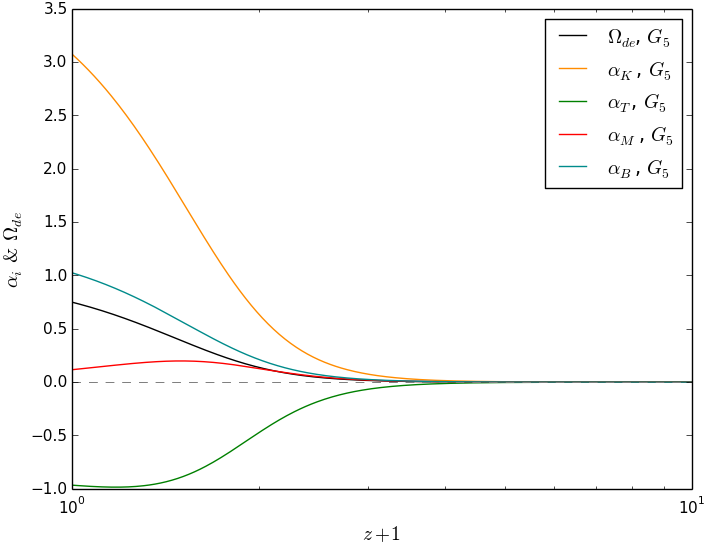}} \\
  \end{minipage}
  \hfill
 \begin{minipage}{0.49\linewidth}
    \centering
	{\includegraphics[width=\textwidth]
	{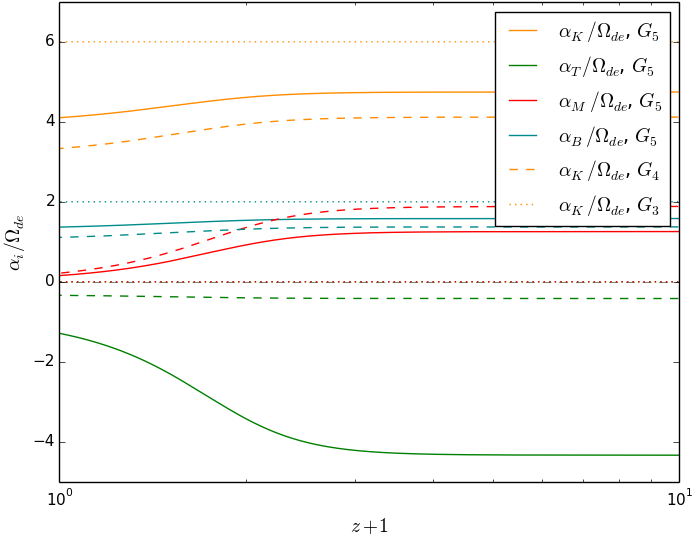}} \\
  \end{minipage}
   \caption{\emph{Left panel}: $\Omega_{de}$ and the $\alpha$-functions for quintic Galileons as a function of redshift. \\
     \emph{Right panel}: The value of the ratio $\alpha_i / \Omega_{de}$ for cubic, quartic and quintic Galileons over the redshift.
     Solid, dashed and dotted lines refer to quintic, quartic and cubic Galileons, respectively.
       Different colors show the evolution of each $\alpha$-function characterizing linear perturbations. The model of the parameterization in which $\alpha_i$ is proportional to $\Omega_{de}$ can reproduce those Galileons models well with nearly constant ratio $\alpha_i / \Omega_{de}$.
     Note that cubic Galileons with massive neutrinos show the same behaviour as cubic Galileons without massive neutrinos.
   }
  \label{fig:alpha_ov_omega}
\end{figure}

To see if the parameterized models we use can reproduce self-consistent gravity models, we test if they can approximate the Covariant Galileons.
The four $\alpha$-functions for the different Galileon models are plotted in the \emph{left panel} of \autoref{fig:alpha_ov_omega}. In the parameterization models we fixed the $\alpha_i$'s to be directly proportional to $\Omega_{de}$ and only vary the proportionality coefficients $c_i$. Therefore the restriction to this models is good if the ratio $\alpha_i / \Omega_{de}$ is approximately constant in time for Galileons.
This ratio is shown in the \emph{right panel} of \autoref{fig:alpha_ov_omega}. The parameterization $\alpha_B=c_B \Omega_{de}$ is exact for cubic, shift-symmetric theories in the attractor.
We have checked the qualitative behaviour of the perturbations, the GNC's and their modifications w.r.t. $\Lambda$-GR for cubic Galileons and they are perfectly consistent, despite the different evolution of the equation of state in Galileons.

\section{Galaxy number counts and relativistic effects in Horndeski gravity}
\label{sec:GNC}

Let us now analyse how modifications of gravity in the framework of Horndeski theories modify the galaxy number count power spectrum, including all relativistic effects. As an example for fully self-consistent gravity models we consider Covariant Galileons, more precisely cubic Galileons with and without massive neutrinos, as well as quartic and quintic Galileons.
After that we investigate how the different $\alpha$-functions of the effective field theory formulation of Horndeski's gravity affect the galaxy number counts. For each model we analyse how the spectrum of same and different redshift bin correlations is changed w.r.t. a standard General Relativity (GR) model including a cosmological constant, $\Lambda$-GR, and to what extent each of the relativistic corrections contributes to this difference. By looking into the growth of perturbations in the different models and how they are related to the relativistic effects we describe the origin of these deviations and their relation to the properties of gravity.

In the following we consider for the Galileon models the best fit background values from Barreira et al. \cite{Barreira:2014ija}. For the massive neutrinos in cubic Galileons the masses are given by $\Omega_\nu h^2 = 3\times0.193\cdot10^{-2}$ (degenerate mass for the three families) \cite{Barreira:2014ija}.
For all parameterized Horndeski models we use cosmological parameters consistent with Planck \cite{Ade:2015xua}.
The power spectra are considered within tophat redshift bins of width $\Delta z=0.05$, which represents an intermediate configuration between those typically reachable by photometric and spectroscopic redshift determinations.
To test cosmological scales in both the early and late Dark Energy era, we consider the redshift bins $z=0.3$ and $z = 2.5$ as well as the intermediate bin $z=1$.
We set the galaxy bias $b$ to one, neglect magnification and evolution biases, $s=f_{\rm evo}=0$, and assume a flat galaxy distribution.
All these effects would be relevant for a consistent detection analysis but do not affect our comparison of gravity models.

\begin{figure}[t]
 \centering
   {\includegraphics[width=0.49\textwidth]
   {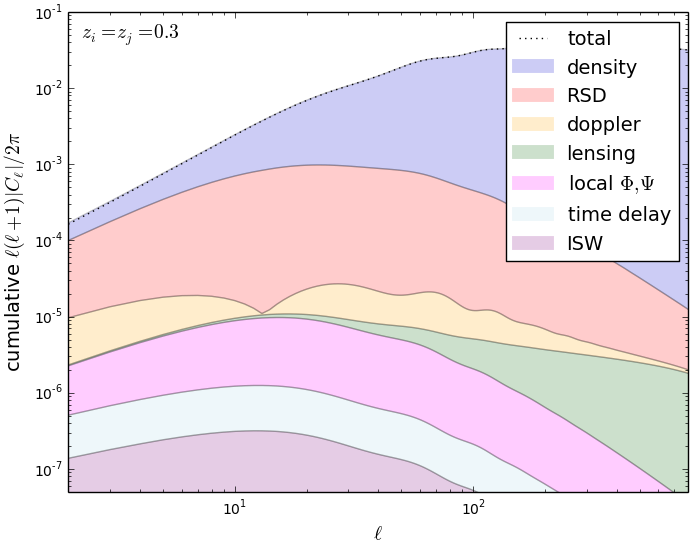}}
   {\includegraphics[width=0.49\textwidth]{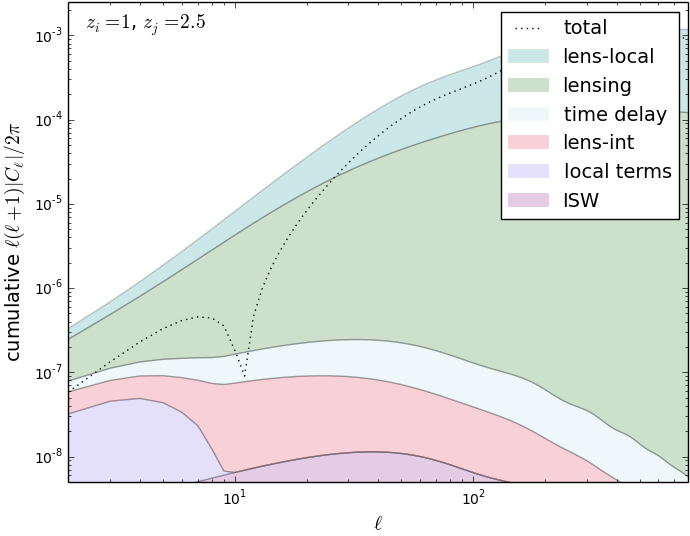}}

   \caption{Galaxy number count angular power spectrum for $\Lambda$-GR where the coloured regions indicate the contribution of each relativistic effect (\ref{eq:D_delta}-\ref{eq:D_pot_isw}) to the total signal at $z_i=z_j = 0.3$ (\emph{left}) and $z_i = 1, z_j=2.5$ (\emph{right}). Each line corresponds to the contribution of an effect and its cross-correlation with density, including in sequence the dominant terms as explained in the text. We used a tophat window function with bin width $\Delta = 0.05$. Note that the total signal in different bin correlations (\emph{right panel}) is negative for $\ell \gtrsim 10$, as the dominant local-lensing cross-correlation is negative. The lensing contribution is positive, therefore the total signal is smaller than the sum of the parts in absolute value.
 \label{fig:CommulEff}   }
\end{figure}

Before discussing our results for modified gravity theories we will give a brief summary of the most important characteristics of the galaxy number count power spectrum for $\Lambda$-GR. We show the total GNC's in \autoref{fig:CommulEff} for same (\emph{left panel}) and different (\emph{right panel}) redshift bin correlations.
Each line corresponds to the contribution of an effect and its cross-correlation with density, including in sequence the dominant terms. For example the purple line and shaded region represent the contribution from the ISW effect. The blue line from the time-delay corresponds to the sum of time-delay and ISW effect while the blue area indicates the contribution of the time-delay to it.
In the case where only one redshift bin is considered the total GNC's are dominated by the intrinsic density fluctuations and RSD. The lensing convergence and its cross-correlation with the density term play a minor role on ultra-large scales but can, depending on the configuration, overcome the RSD on smaller scales (e.g. $\ell \gtrsim 120$ for $z_i=z_j=0.3$). Further contributions, namely Doppler effects, local potential terms, Shapiro time-delay and the ISW effect, are sub-dominant. For more details see \cite{Bonvin:2011bg}.

In the correlation of two different redshift bins (\emph{right panel} of \autoref{fig:CommulEff}) the importance of the local effects decreases with further separation of the bins. The signal is dominated by the negative cross-correlation of lensing convergence and the local effects \cite{Montanari:2015rga}. The pure lensing contribution is the second largest effect followed by the Shapiro time-delay and ISW effect, if considered together with their cross-correlation with the density fluctuations. For redshift bins which are closer together or overlap, local contributions can become important on ultra-large scales.

In addition, the cross-correlations between the lensing convergence and the integrated relativistic effects, Shapiro time-delay and ISW effect, are also important to take into account on ultra-large scales in the correlation of high, well separated redshift bins \cite{Raccanelli:2013gja}.
An interesting feature of this redshift bin configuration is that there can be a change of sign in the signal. This happens when the positive lensing convergence is larger than the negative cross-correlation between the local and lensing terms on ultra-large scales. For $\Lambda$-GR this is the case for $\ell$ up to $10$. On the scale where the two major effects cancel each other out the signal is dominated by the Shapiro time-delay. This is shown in the right panel of \autoref{fig:CommulEff}.
We stress that in practice the sign of the correlation will also depend on observational factors, such as the value of the magnification bias term $2-5s(z)$ entering in eq.~(\ref{eq:D_kappa}).

\subsection{Covariant Galileons}
To understand the modifications to the GNC's, we first analyse matter and metric perturbations in the different Galileon models. We use the best fit models for cubic (with and without massive neutrinos), quartic and quintic Galileons from Barreira et al. \cite{Barreira:2014jha}.

All models predict a larger growth of matter perturbations, as discussed in \ref{sec:CovGal}. Furthermore the Bardeen potentials are subject to scale depended modifications, which has already been addressed in \cite{Barreira:2014jha}.
Cubic Galileon models do not incorporate anisotropic stress, therefore both Bardeen potentials are changed in the same manner. For modes close to the horizon the modifications are minor while on sub-horizon scales the potentials increase after matter domination, instead of a decrease as in $\Lambda$-GR.
In quartic Galileons the potentials grow as well in the Dark Energy dominated era for sub-horizon modes but the evolution is very flat compared to $\Lambda$-GR. $\Phi$ can even stay almost constant for modes $k \gtrsim 100 \mathcal{H}_0$ while $\Psi$ increases up to a turning point and slowly decreases afterwards.
For quintic Galileons the potential increase monotonically in the Dark Energy dominated era for sub-horizon modes while they decrease in $\Lambda$-GR. The deviation w.r.t. $\Lambda$-GR on sub-horizon scales gradually decreases for increasingly larger $k$.
The modification in the case of cubic Galileons is larger as the extra terms in quartic and quintic models reduce the influence of the fifth force which lead to less rapid changes in the evolution of the Bardeen potentials w.r.t. the cubic case \cite{Barreira:2014jha}.

\begin{figure}[t!]
 \centering
  \begin{minipage}{0.49\linewidth}
    \centering
    \subfloat [auto \& cross-correlations]
	{\includegraphics[width=\textwidth]
	{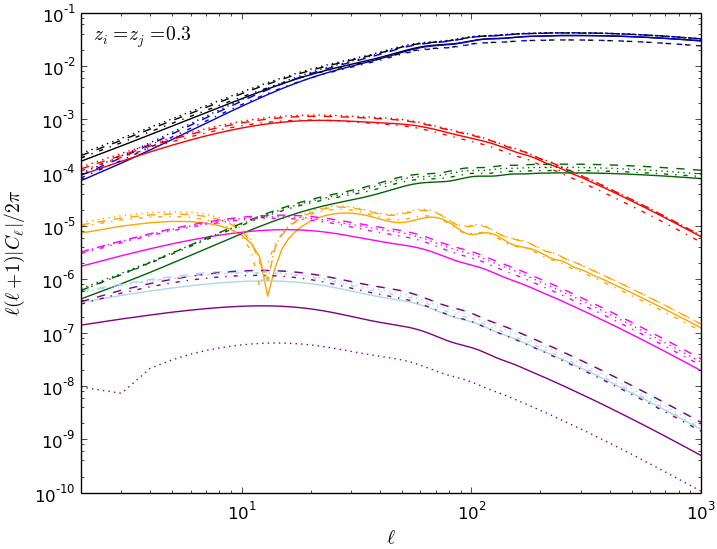}} \\
	{\includegraphics[width=\textwidth]{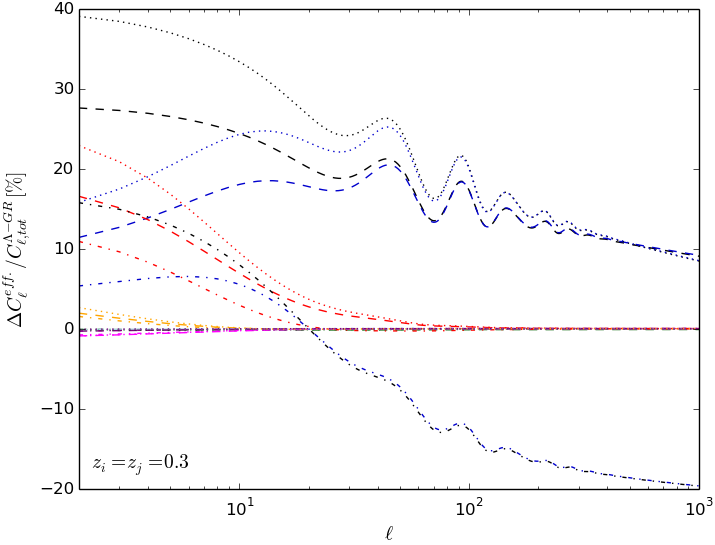}}
  \end{minipage}
  \hfill
 \begin{minipage}{0.49\linewidth}
    \centering
 	\subfloat[auto correlations only]
	{\includegraphics[width=\textwidth]
   {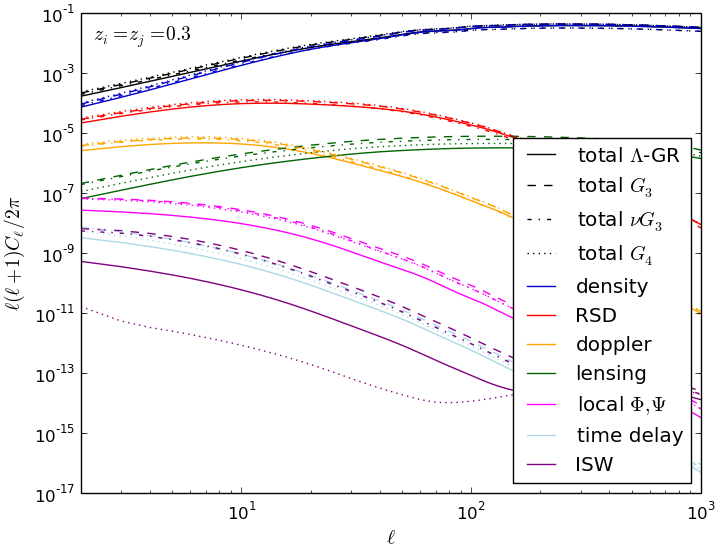}} \\
   {\includegraphics[width=\textwidth]			  {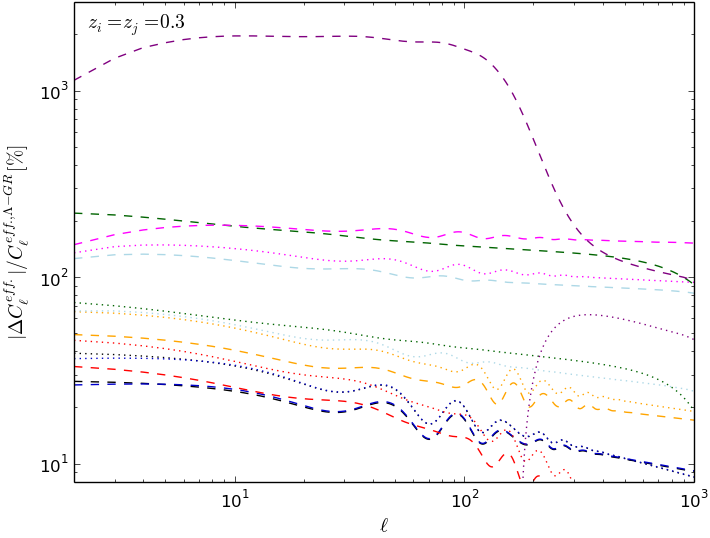}}
  \end{minipage}
 \caption{Relativistic effects and total galaxy number count power spectrum for Galileon models and $\Lambda$-GR for $z_i=z_j = 0.3$ and tophat window function with bin width $0.05$. \emph{Left}: The relativistic effects including their cross-correlation with density is shown. The fractional deviations w.r.t. the total deviation from $\Lambda$-GR are plotted in the lower panel. \emph{Right}: Pure auto correlations of each effect and the fractional deviation w.r.t. the corresponding effect $\Lambda$-GR. Note that we have omitted $\nu G_3$ in the fractional deviations for the purpose of distinctiveness. The variations tend to be smaller than in $G_3$ but of the same order of magnitude.
  For a better visualization of the lower panels, we use a linear scale on the left and a logarithmic one on the right.
  \label{fig:GalAutoBin}   }
\end{figure}

The enhanced growth of the Bardeen potentials and matter perturbations lead to an increase of the intrinsic density fluctuations, the velocity terms and all contributions which directly depend on the potentials. The total GNC's can be enhanced up to 40\% w.r.t. $\Lambda$-GR on ultra-large scales.
We recall that we fix the amplitude of the CMB fluctuations, already well constrained \cite{Ade:2015rim}. We show the total GNC's and all relativistic contributions to it in \autoref{fig:GalAutoBin} for $\Lambda$-GR, cubic, $G_3$, cubic neutrino, $\nu G_3$, and quartic Galileons, $G_4$, for $z = 0.3$.
Each effect with its cross-correlation with density is shown in the \emph{upper left panel}. The \emph{lower left panel} shows the fractional deviation of each effect w.r.t. the total GNC's of $\Lambda$-GR. It is computed as
  \begin{equation} \label{eq:dev}
    \frac{C_{\ell}^{eff.}-C_{\ell}^{eff.,\;\Lambda-GR}}{C_{\ell}^{\Lambda-GR}} \;,
  \end{equation}
  where in the numerator only the contribution from a single effect (but including its correlations with the density) is considered, while in the denominator we consider the total spectra with all contributions.
To see how the single effects are changed more clearly, we plot only the pure effects and their fractional deviation w.r.t. the corresponding effect in $\Lambda$-GR in the \emph{right panels}. The oscillations which are visible in the plots for scales $\ell \gtrsim 30$ are due to the misalignment of BAOs in Galileon models caused by the different expansion histories.
Indeed, given the BAOs scale $r_{\rm BAO}\approx110\,{\rm Mpc}/h$, we expect $\ell_{\rm BAO}=d(z)\frac{2\pi}{r_{\rm BAO}}\approx50$ with the comoving distance $d(z)$ at $z=0.3$.

We find that the integrated effects are the ones which are most sensitive to the changes of gravity in the considered models. For the lensing convergence and the Shapiro time-delay the enhancement w.r.t. $\Lambda$-GR is of order $100\%$, and it can be up to $2000\%$ for the ISW effect.
Note that it has a different sign in Galileon models caused by the changed slope of the Bardeen potentials in comparison to $\Lambda$-GR.
While the negative sign of the ISW effect in Galileons is not visible in the auto correlation $\left\langle \Delta_{ISW} \Delta_{ISW} \right\rangle$ (right panel of \autoref{fig:GalAutoBin}), it becomes relevant in the cross-correlation with density, $\left\langle \Delta_{\delta} \Delta_{ISW} \right\rangle$. Therefore the contribution from the ISW effect including its cross-correlation with density is negative in contrast to $\Lambda$-GR (note that we plot absolute values in the upper left panel of \autoref{fig:GalAutoBin}). This signature will be further discussed in \autoref{sec:CMB-LSS}.

Even though the changes in the ISW effect can be very large, it is the least dominant contributions to the total GNC's and will therefore be hard to detect even in the most optimistic cases. A summary of our results is given in \autoref{tab:Gal}. It shows the contribution of each effect to the deviation of the $C_\ell(z_i, z_i)$'s with respect to $\Lambda$-GR for the fixed scale $\ell = 10$ for the redshift bins $z_i = 0.3,1$ and $2.5$. We also include quintic Galileons, $G_5$, which we have omitted from \autoref{fig:GalAutoBin} for the purpose of clearness. The effects show a similar scale dependence as in quartic Galileons. The modifications to local terms are less drastic while the terms integrated along the line of sight are increased as the metric perturbation in the quintic case are more enhanced.

\begin{table}[t] \makebox[\linewidth]{
 \begin{tabular}{l | c | c | c c | c |cc c}
Model & Total & Density & RSD & Doppler & Lensing & Lpot & t-delay & ISW \\ \hline  \hline

$G_3$&24&18 (25) &7 (26) & -  (40) &-0.2 (187) &-0.3 (189) & -  (123) & -  (1964)  \\
 $G_3$&4&-0.4 (-1.1) &3 (9) & -  (19) &1.0 (132) & -  (84) & -  (38) & -  (835)  \\
 $G_3$&1.1&-2 (-6) &-2 (-2) & -  (-11) &5 (99) & -  (23) & -  (11) & -  (441)  \\
  \hline$\nu G_3$&8&6 (8) &3 (14) & -  (28) &-0.1 (145) &-0.2 (165) & -  (105) & -  (1460)  \\
 $\nu G_3$&-5&-4 (-10) &-2 (1.1) & -  (8) &0.8 (110) & -  (67) & -  (39) & -  (563)  \\
 $\nu G_3$&-8&-4 (-15) &-9 (-12) & -  (-17) &4 (84) & -  (163) & -  (16) & -  (287)  \\
  \hline$G_4$&33&24 (34) &9 (35) & -  (53) &-0.2 (59) &-0.3 (142) & -  (56) & -  (-99)  \\
 $G_4$&8&1.2 (3) &7 (16) & -  (30) &0.5 (64) & -  (136) & -  (36) & -  (-19)  \\
 $G_4$&0.6&-1.2 (-4) &-1.0 (0.2) & -  (-7) &3 (56) & -  (-22) & -  (14) & -  (-0.3)  \\
  \hline$G_5$&28&21 (30) &7 (21) & -  (37) &-0.2 (91) &-0.2 (109) & -  (74) & -  (103)  \\
 $G_5$&6&1.0 (3) &5 (11) & -  (24) &0.6 (81) & -  (79) & -  (32) & -  (147)  \\
 $G_5$&2&-1.0 (-4) &-0.5 (0.8) & -  (-7) &3 (66) & -  (5) & -  (11) & -  (50)  \\
  \hline
  \end{tabular}}
  \caption{Contribution of each effect including its cross-correlation with density to the deviation of the total GNC's (computed according to eq.~(\ref{eq:dev})) in percentage. Numbers in parenthesis give the relative deviation of each individual effect w.r.t. the corresponding effect in $\Lambda$-GR. First, second and third line represent the values for the redshift bins $z_i=z_j=0.3,1$ and $2.5$, respectively. All departures are computed at $\ell=10$; `` - '' indicates that the absolute value of a deviation is smaller than 0.1\%.
    \label{tab:Gal}}
\end{table}

\begin{figure}[t]
 \centering
  \begin{minipage}{0.49\linewidth}
    \centering
   {\includegraphics[width=\textwidth]
   {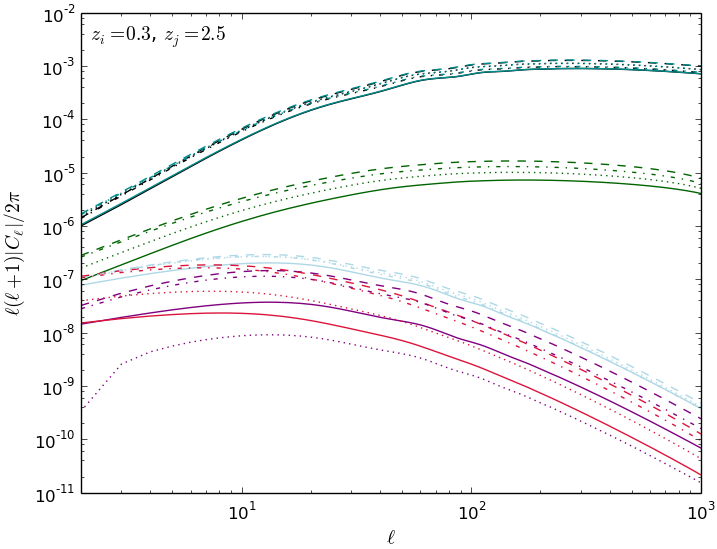}} \\
  \end{minipage}
  \hfill
 \begin{minipage}{0.49\linewidth}
    \centering
	{\includegraphics[width=\textwidth]
	{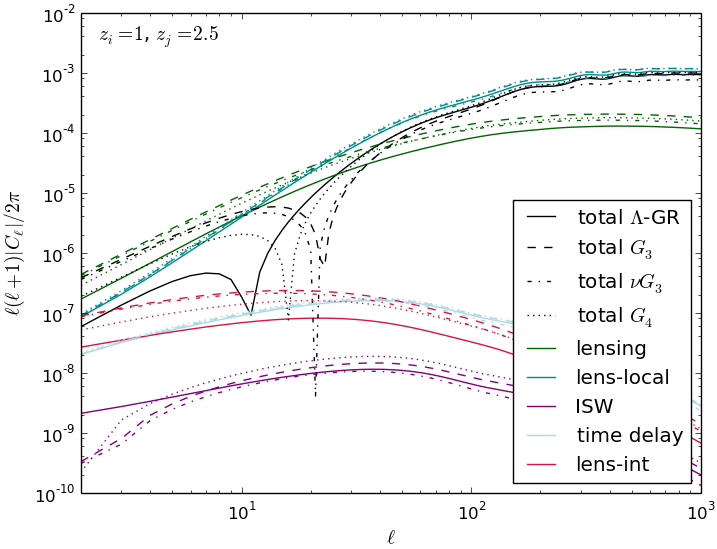}} \\
  \end{minipage}
 \caption{Dominant relativistic effects and total galaxy number count power spectrum for Galileon models and $\Lambda$-GR in the correlation of different redshift bins. Time-delay and the ISW effect include their cross-correlation with density. The window function is a tophat with bin width $0.05$.
   The redshift correlations are indicated in the plots. Correlations of the lensing term alone can dominate the signal on the largest angular scales.
  \label{fig:GalCrossBin} }
\end{figure}

In configurations with two different redshift bins the changes are more drastic as the dominating contributions depend on the lensing convergence. Due to the enhancement of the lensing potential in Galileon models the total signal is increased of order 50\% even on scales up to $\ell \sim 100$. \autoref{fig:GalCrossBin} shows the GNC's and the most important contributions for two different redshift bin configurations.
Looking at the \emph{right panel} of \autoref{fig:GalCrossBin}, the scale on which the total signal changes its sign for $z_i=1,z_j=2.5$ is shifted towards smaller scales (larger $\ell$) for all Galileon models. This is due to the fact that local terms are not increased as dramatically as integrated terms and hence the scale on which they are canceled out by lensing is decreased.
It is also interesting to notice that in Galileon models the cross-correlation between lensing and the integrated relativistic effects can become larger than the total GNC's in $\Lambda$-GR on ultra-large scales.
The fractional deviations to $\Lambda$-GR for different redshift bin correlations on the fixed scale $\ell = 10$ are displayed in \autoref{tab:GalCrossBin}.

\begin{table}[t] \makebox[\linewidth]{
 \begin{tabular}{l | c | c | cc c | cc}
Model & $z_i - z_j$ & Total & Lensing & Lensing-Local & Lensing-Int. & t-delay & ISW \\ \hline  \hline
$G_3$ & 0.3-1&61&58 (175) &65 (59) &-0.5 (185) &0.5 (102) &2 (1831)  \\
 $G_3$ & 0.3-2.5&58&57 (171) &63 (58) &-0.4 (181) &0.2 (97) &0.4 (1619)  \\
  \hline$\nu G_3$ & 0.3-1&45&43 (138) &48 (44) &-0.4 (147) &0.4 (90) &1.4 (1323)  \\
 $\nu G_3$ & 0.3-2.5&43&43 (136) &47 (43) &-0.3 (144) &0.2 (88) &0.4 (1173)  \\
  \hline$G_4$ & 0.3-1&51&50 (60) &53 (51) & -  (62) &0.4 (46) &0.4 (-98)  \\
 $G_4$ & 0.3-2.5&49&49 (60) &51 (49) & -  (61) &0.2 (43) &0.1 (-98)  \\
  \hline$G_5$ & 0.3-1&53&52 (90) &55 (53) &-0.2 (94) &0.4 (62) &0.9 (115)  \\
 $G_5$ & 0.3-2.5&51&51 (89) &53 (51) &-0.2 (92) &0.2 (58) &0.2 (91)  \\
  \hline
  \end{tabular} }
  \caption{Contribution of the dominating effects in different redshift bin correlations to the deviation of the total GNC's (computed according to eq.~(\ref{eq:dev})) in percentage. The Shapiro time-delay and the ISW effect include their cross-correlation with density. Numbers in parenthesis give the relative deviation of each individual effect w.r.t. the corresponding effect in $\Lambda$-GR. All departures are computed at $\ell=10$; `` - '' indicates that the absolute value of a deviation is smaller than 0.1\%. Note that we have omitted $z_i = 1$, $z_j= 2.5$ as the fractional deviations diverge due to the zero crossing of the signal in $\Lambda$-GR on this scale.
  \label{tab:GalCrossBin}}
\end{table}

The differences in the GNC's between the Galileon models can be understood as follows: adding massive neutrinos to a model introduces the usual suppression of power on small scales due to their free streaming, as it is shown for a cubic Galileon example. This effect partly compensates the enhanced growth.
Quartic and quintic Galileons have a higher expansion rate than cubic models leading to increased growth in the matter era. Therefore the dominant local terms, i.e. intrinsic density fluctuations, RSD and Doppler terms, are enhanced leading to an increase of the total signal. However, the other sub-dominant effects which depend directly on the Bardeen potentials are less modified than in cubic Galileons as the extra terms reduce the power of the fifth force and therefore the changes to the metric perturbations.

\subsection{Parameterized Horndeski gravity}
\label{sec:Param}
We consider now the effects of the different $\alpha$-functions from the Bellini-Sawicki parameterization \cite{Bellini:2014fua} on each contribution to the GNC's. For this purpose we use models in which the $\alpha$-functions are proportional to the energy density of Dark Energy, i.e. $\alpha_i = c_i \Omega_{de}$, where the $\alpha_i$'s and $c_i$'s correspond to kineticity, braiding, running and tensor speed excess respectively. To make sure we are considering viable theories we run several Markov Chain Monte Carlo analyses using Planck 2015 \cite{Ade:2015xua} temperature data and BAO measurements from BOSS \cite{Aubourg:2014yra,Font-Ribera:2013wce,Delubac:2014aqe}. Our MCMC analysis showed that the standard cosmological parameters do not vary significantly from the $\Lambda$GR values. We have therefore fixed them to the Planck best-fit values \cite{Ade:2013zuv} when plotting the model predictions. This analysis gives us a baseline of models that produce acceptable CMB spectra and expansion rates but without introducing any information about the matter distribution other than inferred from secondary CMB effects.
As we analyse the effects of either one parameter or of a combination of parameters, we run a MCMC for each of the considered combinations. Our results are shown in \autoref{tab:MCMC}.

\begin{table}
\begin{center}
 \begin{tabular}{c |c| c |c| c  }
 $c_K$ & $c_B$ & $c_M$ & $c_T$ &  $w$ \\ \hline \hline
  1 & $0.63_{-0.47}^{+0.29}$ & 0 & 0 &   $-1$ \\[2pt]
  1 & $0.92_{-0.59}^{+0.40}$ & $1.12_{-1.46}^{+0.74}$ & 0  & $-1$ \\[2pt] \hline
  1 & 0 & $0.44_{-0.44}^{+0.12}$ & 0 &   $-1$ \\[2pt]
  1 & 0 & $0.61_{-0.67}^{+0.19}$ & $0.06_{-1.04}^{+0.33}$  & $-1$ \\ [2pt] \hline
  1 & 0 & 0			& $1/\Omega_{de}<c_T<0$ & $-1$ \\ [2pt]
   \hline \hline
\end{tabular}
\end{center}
\caption{Constraints on the Horndeski parameterization (\ref{eq:propto_omega}) and (\ref{eq:expansion}) using Planck and BOSS-BAO data. Values show the mean and 68\% confidence intervals (see also \cite{Bellini:2015xja}). When errors are not indicated, we fixed the parameter in the analysis. For $c_T$ we have taken the constraints arising purely from stability conditions.
 \label{tab:MCMC}}
\end{table}

\subsubsection{Kineticity}
The kineticity parameterizes the standard kinetic term for the scalar field. It modulates its speed of sound, but does not directly affect the growth of matter or metric perturbations in any significant way. Hence, we do not expect the GNC's to be sensitive to a variation of $c_K$. This expectation was corroborated by our results: even for extreme values like $c_K= 10^{\pm 3}$ the fractional deviations w.r.t $\Lambda$-GR do not exceed 0.1\% on any scale up to $\ell = 1000$ in the $C_\ell(z_i,z_j)$'s for the considered redshifts.

We have also checked if the influence of kineticity increases when one considers models with an equation of state parameter $ w \neq -1$. This allows kineticity models to have a non-zero speed of sound, which is suppressed by $(1+w)$ in this simple case. Considering a value $w=-0.87$ ($2\sigma$ away from $w=-1$ \cite{Ade:2015xua}) and comparing two models with and without $c_K$, does not give rise to any significant changes in the GNC's or the relativistic corrections to it.
For our further analysis we have fixed the value of $c_K$ to 1 in all considered models.

\subsubsection{Tensor Speed Excess}
By tuning the tensor speed excess, $c_T$, one can consider models in which tensor perturbations propagate faster ($c_T>0$) or slower ($c_T<0$) than the speed of light. Models with negative values of $c_T$ lead to a slight reduction of power in the matter power spectrum for modes $k \lesssim 10^{-3} h/Mpc$. Modifications to the metric perturbations also only start becoming relevant on scales close to the horizon.
On these scales the tensor speed excess introduces anisotropic stress and a negative value of $c_T$ causes an increase of $\Phi$ while $\Psi$ is decreased w.r.t. $\Lambda$-GR in the Dark Energy dominated era. The effect on $\Psi$ is larger than the effect on $\Phi$ such that the lensing potential $\propto \Phi + \Psi$, is reduced.

The decrease of matter perturbations and of the lensing potential are reflected in the GNC's. The dominating local effects, the lensing convergence and the time-delay are slightly decreased. Only the ISW effect is decrease for small redshifts while it is enhanced for high redshifts.
However, the modifications to the total signal are very small (below 1\%) on all scales up to $\ell = 1000$ for all considered redshift bin correlations $z_i = z_j$ and $z_i \neq z_j$, even for large values of the tensor speed excess ($c_T = -0.9$).

\subsubsection{Running effective Planck Mass}
\label{sec:Param_Mass}
In this section we first consider the influence of a running of the effective Planck mass (``running'') on the GNC's and then models with both, running and tensor speed excess.

A positive running rate causes an effective increase of the gravitational constant at late time, implying more clustering. Therefore the matter power spectrum is enhanced on ultra-large scales ($k \lesssim 10^{-3} h/Mpc$). Considering the metric perturbations the running causes $\Phi$ to decrease and $\Psi$ to increase on sub-horizon scales, where the effect on $\Phi$ is stronger such that the lensing potential is decreased. The evolution of both potentials is faster than in $\Lambda$-GR leading to a stronger tilt of the curves.

\begin{figure}
 \centering
  \begin{minipage}{0.49\linewidth}
    \centering
    \subfloat [auto \& cross-correlations]
	{\includegraphics[width=\textwidth]{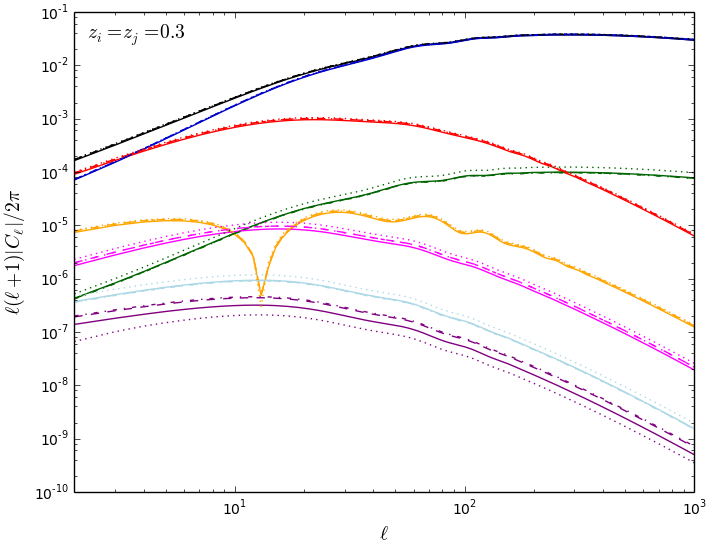}} \\
	{\includegraphics[width=\textwidth]{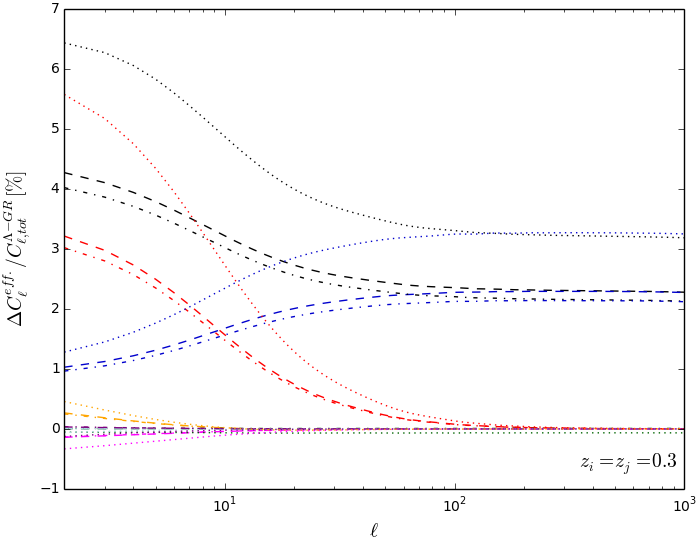}}
  \end{minipage}
  \hfill
 \begin{minipage}{0.49\linewidth}
    \centering
 	\subfloat[auto correlations only]
	{\includegraphics[width=\textwidth]{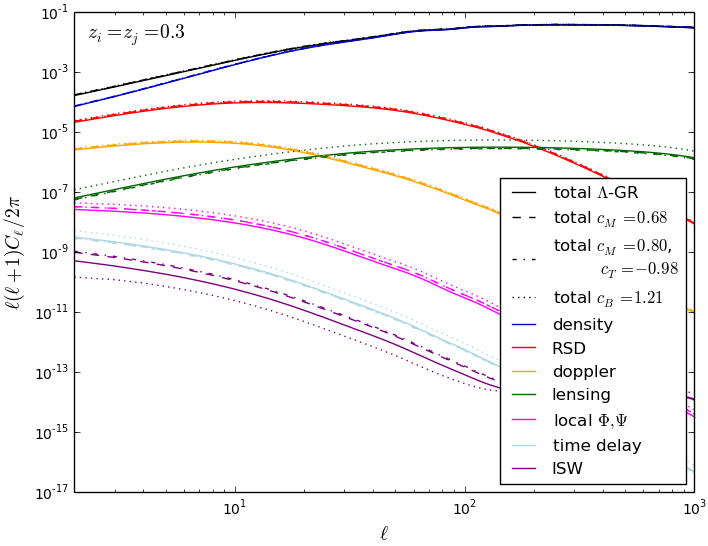}} \\
   {\includegraphics[width=\textwidth]{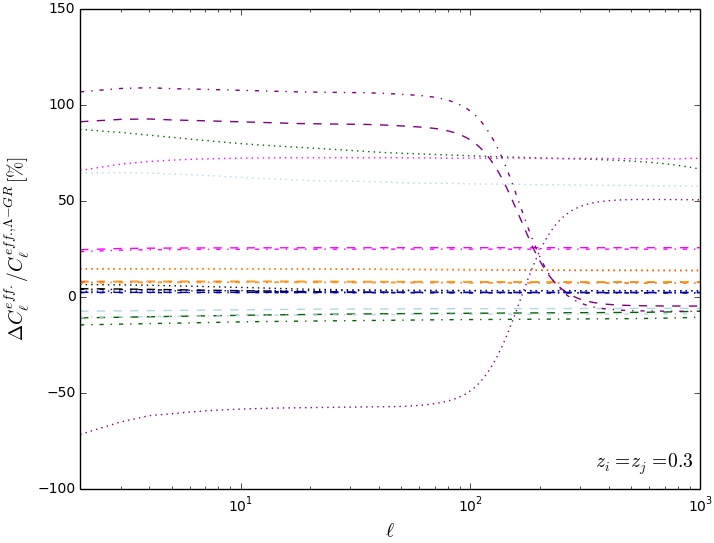}}
  \end{minipage}
 \caption{Relativistic effects and total galaxy number count power spectrum for different parameterized models and $\Lambda$-GR for $z_i=z_j = 0.3$. \emph{Left}: The relativistic effects including their cross-correlation with density is shown. The fractional deviations w.r.t. the total deviation from $\Lambda$-GR are plotted in the lower panel. \emph{Right}: pure auto correlations of each effect and their fractional deviation w.r.t. the corresponding effect $\Lambda$-GR. In each case we used a tophat window function with bin width $\Delta = 0.05$.
	\label{fig:ParAutoBin}}
\end{figure}

In the GNC's the increase of the matter power translates to an enhancement of the density fluctuations, RSD and Doppler effects. On the other hand, the decreased lensing potential leads to and reduction of the lensing convergence and the Shapiro time-delay contributions w.r.t. $\Lambda$-GR. Through the larger tilt of the Bardeen potentials the ISW effect, which depends on their derivatives, gets enhanced.
We show the total GNC's and all relativistic contributions to it for $z = 0.3$ in \autoref{fig:ParAutoBin} for $\Lambda$-GR and a model with $c_M = 0.68$, which corresponds to the mean of our MCMC plus 2$\sigma$. In the \emph{upper left panel} we show each effect including its cross-correlation with density. The \emph{lower left} shows the fractional deviation of each effect w.r.t. the total GNC's of $\Lambda$-GR. The pure effects and their fractional deviation w.r.t. to the corresponding effect in $\Lambda$-GR are displayed in the \emph{right panel}.
It shows that the ISW effect is the most sensitive effect to a variation of $c_M$. But even though the integrated effects can be modified significantly, the effect on the total GNC's $C_\ell(z_i,z_i)$ is only small (\emph{lower left} panel).

\begin{figure}
 \centering
  \begin{minipage}{0.49\linewidth}
    \centering
   {\includegraphics[width=\textwidth]
   {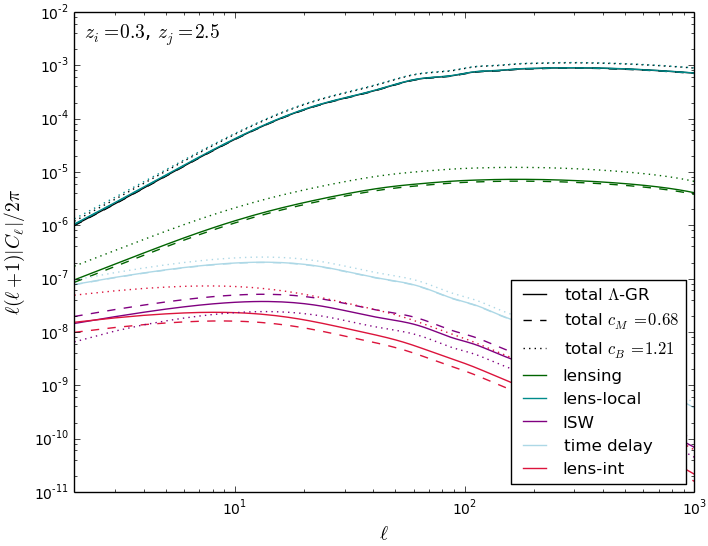}} \\
  \end{minipage}
  \hfill
 \begin{minipage}{0.49\linewidth}
    \centering
	{\includegraphics[width=\textwidth]
	{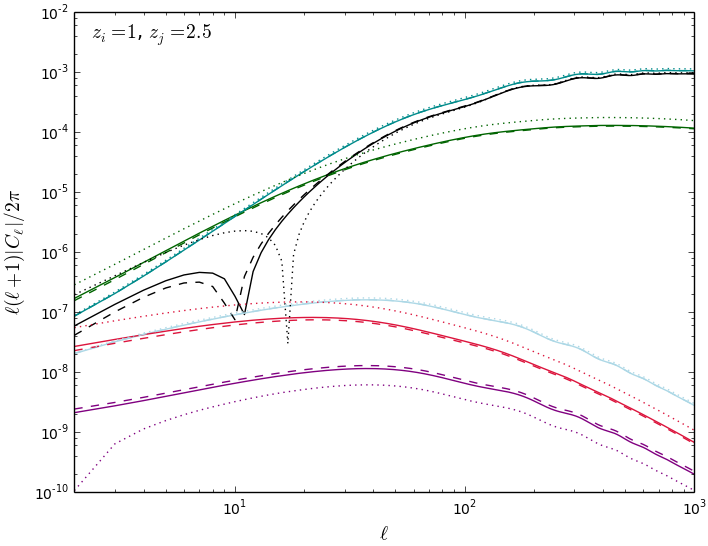}} \\
  \end{minipage}
 \caption{Dominant relativistic effects and total galaxy number count power spectrum for parameterized Horndeski models and $\Lambda$-GR for different redshift bin correlations. The Shapiro time-delay and the ISW effect include their cross-correlation with density. The window function is a tophat with bin width $0.05$. The redshift bins are indicated in the plots. Note that in the case of $z_i=1$,$z_j=2.5$ the signal can even be dominated by the lensing correlations alone on the largest scales.
  \label{fig:ParCrossBin} }
\end{figure}

In different redshift bin correlations the enhanced impact of the lensing convergence causes a stronger modification of the total signal, see \autoref{fig:ParCrossBin}. It it is about 20 \% on all scales up to $\ell = 1000$ for $c_M = 0.68$ in the configuration $z_i=0.3$, $z_j = 2.5$ (\emph{left panel} of \autoref{fig:ParCrossBin}). The \emph{right panel} shows the contributions to GNC's for for $z_i = 1$, $z_j = 2.5$. Here the slight enhancement of the local terms and the decrease of the lensing convergence shifts the scale on which both effects cancel each other out. The sign change of the total signal is therefore shifted towards smaller scales (larger $\ell$) for larger values of $c_M$.

Models with a negative run rate of the effective Planck mass have instabilities associated to a negative speed of sound (see eq.~\ref{eq:c_s}), leading to diverging values of the scalar field on the scales we are considering.
Therefore it is necessary to introduce - at least - one other non zero component in addition to $c_K$ and $c_M$. In this light we also analyse how a combination of the running effective Planck mass and the tensor speed excess affect the perturbations and the GNC's.

In the configuration with $c_M\neq 0$ an introduction of a negative tensor speed excess has the same influence on the perturbations as before for $c_M = 0$ for modes well below the horizon ($k \gtrsim 10\HH_0$). The reduction of the matter power and the lensing potential lead to a decrease of the leading local contributions and the lensing potential. This can be seen very well in the lower panels of \autoref{fig:ParAutoBin} when comparing the two models with $c_M$: as discussed, an increase of $c_M$ leads to an increase of the density, RSD and Doppler contributions, but with tensor speed excess these effects are decreased even w.r.t. a model with smaller $c_M$ but $c_T = 0$.%
\footnote{The model $c_M = 0.9$ and $c_T= -0.98$ corresponds to the mean $c_M + 1 \sigma$ and mean $c_T - 1\sigma$ of our MCMCs. For each parameter we have taken the bound which is further away from standard gravity, i.e. $c_i = 0$, and 1$\sigma$ values to avoid tensor inabilities due to $1 + \alpha_T < 0$.}

For the ISW effect the behaviour is more complex. Modifications of $c_T$ have a larger impact for higher redshifts which is opposite to the effects of $c_M$, see the first two models in \autoref{tab:ISW}.
In combination the presence of the running boosts the impact of the tensor speed excess to become larger for high redshifts. Instead of being decreased for a model with only running, the ISW effect in combination with its cross-correlation with density gets increased for high redshifts by the tensor speed excess, see third model in \autoref{tab:ISW}. This opens a window to potentially distinguish between models with and without $c_T$.

Besides that, it is interesting to notice that considering small modes for a fixed, positive $c_M$, $\Psi$ gets decreased and $\Phi$ increased independently of the sign of $c_T$. However, on modes near the horizon this behaviour is interchanged for negative values of $c_T$ ($\Psi$ is increased while $\Psi$ is decreased). Furthermore, fixing $c_M$, the modifications on the GNC's are stronger for models with tensor modes propagating faster than the speed of light compared to models in which they propagate subluminal.

\begin{table}[t] \begin{center}
\begin{tabular}{l | c | c | c c | ccc cc}
Model & $z_i = z_j $ & ISW & ISW + Dens-ISW  \\ \hline \hline
	$c_T = -0.98$ & 0.3& -0.05  &  -0.08  \\
	$c_T = -0.98$ & 1&-0.2 & -0.1 \\
 $c_T = -0.98$ & 2.5&0.3 & 4 \\
  \hline$c_M = 0.8$ & 0.3&112 & 45 \\
 $c_M = 0.8$ & 1&76 & 18 \\
 $c_M = 0.8$ & 2.5&58 & -3 \\
  \hline$c_M = 0.80$,
         $c_T = -0.98$ & 0.3&108 & 44 \\
 $c_M = 0.80$,
         $c_T = -0.98$ & 1&80 & 24 \\
 $c_M = 0.80$,
         $c_T = -0.98$ & 2.5&68 & 14 \\
  \hline \end{tabular} \end{center}
  \caption{Fractional deviations in percent of the ISW effect and its cross-correlation with density in same redshift bin correlations relative to the corresponding effects in $\Lambda$-GR for $\ell = 10$.
  }
  \label{tab:ISW} \end{table}

\subsubsection{Braiding}
\label{sec:Param_Braid}
In contrast to models with a running of the effective Planck mass or tensor speed excess, braiding models do not incorporate anisotropic stress. Both Bardeen potentials get enhanced in the same manner w.r.t. $\Lambda$-GR and can stay constant even after matter domination for values $c_B \sim 0.8$ (and $c_K = 1$). While the metric perturbations are enhanced for all modes the matter perturbations are scale dependent. The matter power spectrum is slightly enhanced on small scales ($k \gtrsim  2 \cdot 10^{-3} h/Mpc$) for positive values of $c_B$ while it is decreased on larger scales.

The enhanced growth of matter perturbations on sub-horizon scales for models with positive braiding leads to an increase of the dominant local terms. Hence, the total signal gets enhanced of about 6\% for the mean $c_B +2 \sigma$ from our MCMCs ($c_B = 1.21$, $c_K = 1$).
With the increase of the metric perturbations the lensing potential, and thereby the lensing convergence and the Shapiro time-delay, are enhanced as well. For higher redshifts the contribution of the lensing convergence to the total signal gets more important and the deviations of the total GNC's can be dominated by the enhancement of lensing. For example in the considered braiding model for $z = 2.5$ the lensing contribution is increased by $\sim50\%$ which leads to an enhancement of the total signal of order 5\%.
The ISW effect gradually decreases with an increase of $c_B$ and can almost vanish for the case were the Bardeen potentials stay approximately constant even after matter domination ($c_B \sim 0.8$). Increasing $c_B$ further leads to a change of sign in the ISW effect as the potential will increase after matter domination instead of decrease as in $\Lambda$-GR. Therefore the ISW effect is the one which is most sensitive to a change of $c_B$.
All relativistic effects and their deviations from $\Lambda$-GR for this model are shown in \autoref{fig:ParAutoBin} for $z = 0.3$.

A variation of braiding also leaves signatures in different redshift bin correlations. Due to the enhancement of the lensing convergence the cross-correlations between lensing and local as well as lensing and integrated effects are also increased.
As these are the dominant contributions this can lead to an increase of the total signal of $\sim 20\%$ on all scales for $c_B = 1.21$. Another effect is that in configurations with a sign change of the total GNC's the signal can be several orders of magnitude larger or smaller w.r.t. $\Lambda$-GR. This is the case as the negative cross-correlation between lensing and the local terms gets less enhanced than the pure lensing convergence contribution. Therefore the scale where this two effects cancel out is shifted towards smaller scales when increasing $c_B$ (and therefore the lensing convergence). For models with negative braiding the lensing potential is decreased and the shift is towards larger scales.
We stress that for the detection of this effect, observational factors, like the magnification bias term $2-5s(z)$ entering in eq.~(\ref{eq:D_kappa}), also influence the sign of the lensing-local cross-correlation and have to be taken into account (we have set $s=0$ here).

In combinations of braiding with tensor speed excess or a running effective Planck mass only the terms which do not directly depend on the Bardeen potentials, namely density fluctuation, RSD and Doppler, show the expected behaviour from the previous sections when varying one parameter while the others are kept constant.
For the other contributions the dependencies are not as trivial anymore. The modifications to the Bardeen potentials strongly depend on the specific values of $c_B$ and $c_T$ or $c_M$ and can not any longer be related to obvious effects, e.g. a change of sign of one parameter. Furthermore the introduction of anisotropic stress with $c_T$ or $c_M$ leads to additional variations. Therefore the modifications to the relativistic effects depending on the Bardeen potentials in the combined cases are more subtle and have to be analysed for the specific cases separately.

\section{Deep-redshift GNC correlations}
\label{sec:BinCrossCorr}
In this section we explore how GNC correlations between different redshifts can be a test of gravity, even if the considered redshifts are high and, for the models we consider, gravity is very close to GR in the corresponding epoch. This is possible because such correlations are dominated by the interplay between integrated effects, which are affected by lower redshifts through the integration along the line of sight, and their cross-correlation with local effects.

\begin{figure} [t]
 \centering
  \begin{minipage}{0.49\linewidth}
    \centering
    {\includegraphics[width=\textwidth]
	{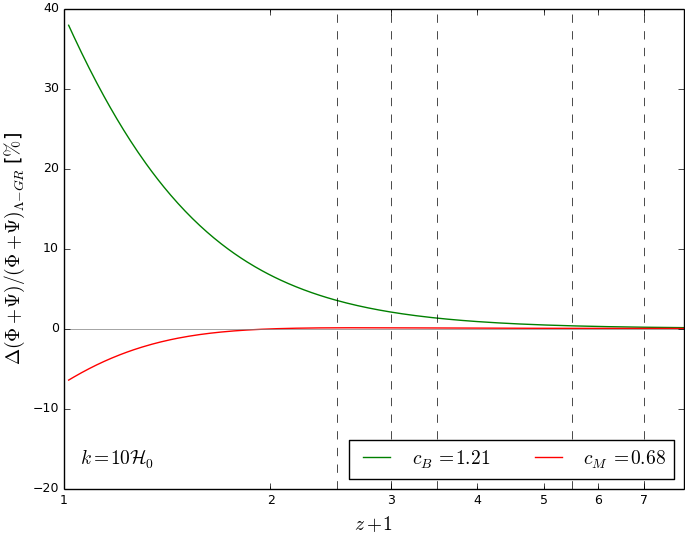}} \\
	{\includegraphics[width=\textwidth]
	{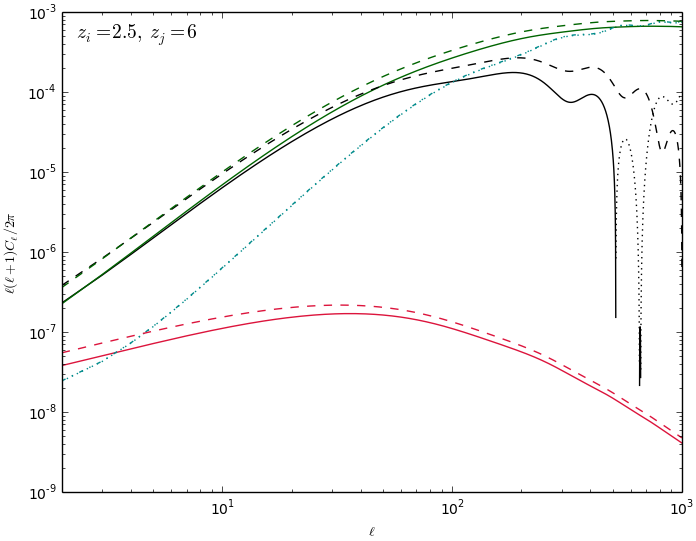}} \\

  \end{minipage}
  \hfill
 \begin{minipage}{0.49\linewidth}
    \centering
 	{\includegraphics[width=\textwidth]
   {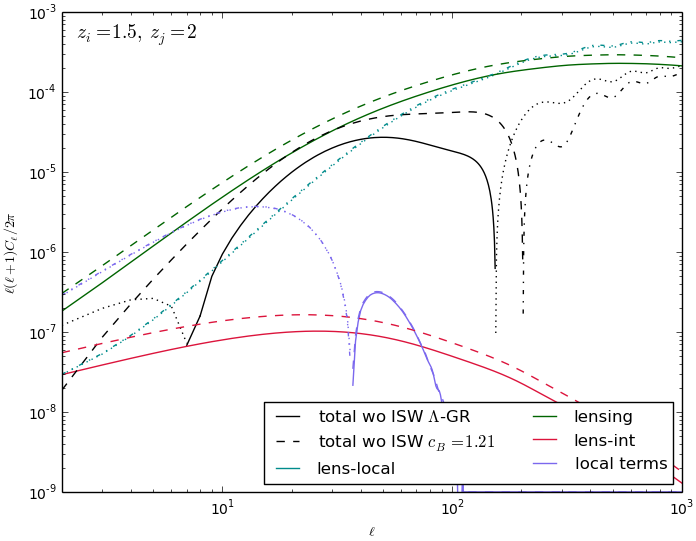}} \\
 	{\includegraphics[width=\textwidth]
	{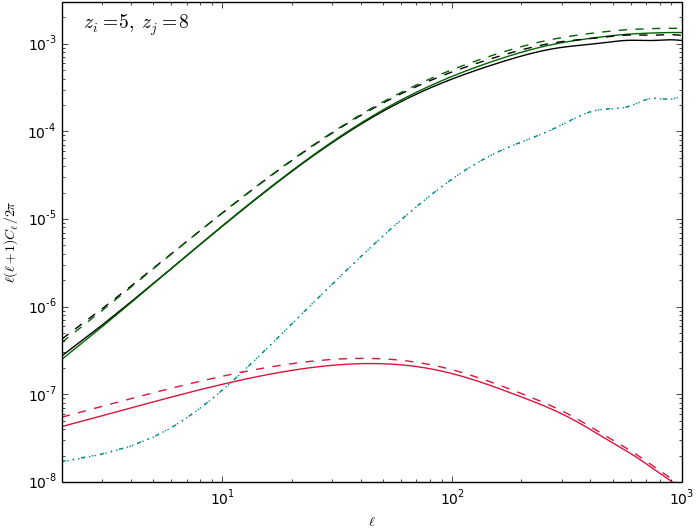}} \\

  \end{minipage}
 \caption{\emph{Upper left:} Fractional deviation of the lensing potential w.r.t. $\Lambda$-GR for a parameterized braiding ($c_B = 1.21, \, c_K = 1$) and a running model ($c_M = 0.86, \, c_K = 1$) in percent for the mode $k = 10 \mathcal{H}_0$. Vertical dashed lines indicate the considered redshift bins ($z = 1.5, 2, 2.5, 4.5$ and $6$). \emph{Upper right and lower panels}: Dominant relativistic effects and total galaxy number count power spectrum for $\Lambda$-GR and the the braiding model for different redshift bin correlations. The window function is a tophat with bin width $0.05$, redshift bins are indicated in the plots. Dotted and dashed-dotted lines indicate negative values for $\Lambda$-GR and braiding respectively. In these plots the contributions from the cross-correlation of local effects with lensing are indistinguishable for the two models. We do not show individual sub-leading terms for clearness.
  \label{fig:HighCrossBin} }
\end{figure}

The lensing convergence contribution to the GNC's, \autoref{eq:D_kappa}, involves an integral of the sum of Bardeen potentials, weighted by a background-dependent factor. As this integral spans from $z=0$ to the redshift of observation, the value of $\Delta_\kappa$ is sensitive to the low redshift evolution of the lensing potential, even when evaluated at a high redshifts. This makes deep redshift GNC cross-correlations sensitive to low $z$ modifications of gravity, even when $z_i,z_j$ are much higher than the typical redshifts on which Dark Energy dominates.
If we compare models that recover GR at high redshifts with the same background expansion, then
\begin{equation}\label{eq:lens_decomp}
(\Delta_\kappa^{\rm MG} - \Delta_\kappa^{\rm GR})[z] =
\frac{\mathcal{R}}{r(z)} + \mathcal{I}\,,
\end{equation}
where both $\mathcal{R},\mathcal{I}$ tend to a constant for $z>z_{\rm GR}$, where $z_{\rm GR}$ is some redshift at which GR is approximately recovered (e.g. top-left panel of \autoref{fig:HighCrossBin}). Note that the last term in (\ref{eq:lens_decomp}) is constant for any $z>z_{\rm GR}$.
The persistence of modifications to the cross-correlations at high redshift is shown in \autoref{fig:HighCrossBin}. We plot the fractional deviations of the lensing potential for two parameterized models with braiding and running of the effective Planck mass (\emph{upper left panel}) and the correlation of different, high redshift bins for three configurations. For comparison with $\Lambda$-GR we use the previously considered braiding model ($c_K = 1,\, c_B = 1.21$) and discuss the effects in a model with a running effective Planck mass ($c_K = 1,\, c_M = 0.68$) qualitatively.

The impact of late time modifications of gravity on the lensing contributions can be used to test gravity with galaxy correlations at different redshifts. Specifically, it leads to several effects on the GNC angular power spectra:
\begin{itemize}
 \item Change of the amplitude of the integrated-integrated contribution (dominated by the lensing term) on the total signal (\emph{upper right} and \emph{bottom panels} of \autoref{fig:HighCrossBin}). This effect can lead to an considerable increase of the value on intermediate scales on which the signal-to-noise is larger.

 \item Shift of the scale at which the spectrum changes sign (\emph{lower left} and \emph{upper right panel} of figure \autoref{fig:HighCrossBin}). Changes in sign come from the interplay of the correlations $\left\langle \Delta_{\rm local} \Delta_{\kappa} \right\rangle$ (negative, dominant on high $\ell$), $\left\langle \Delta_{\kappa} \Delta_{\kappa} \right\rangle$ (positive, important on large scales) and $\left\langle \Delta_{\rm local} \Delta_{\rm local} \right\rangle$ (positive, can be dominant on very low $\ell$ for redshift bins which are nearby). By choosing the redshifts $z_i,z_j$ appropriately, the sign change can be displaced to intermediate angular scales on which the signal to noise is larger (e.g. choosing $z_i=1.5,z_j=2$). Moreover, in this regime the crossing from positive to negative correlation can be shifted significantly in $\ell$ depending on the theory of gravity.

 \item For some redshift combinations this cancellation occurs at the same scale as BAO oscillations (\emph{bottom left panel} of \autoref{fig:HighCrossBin}). Since those are imprinted on the $\left\langle \Delta_{\rm local} \Delta_{\kappa} \right\rangle$ correlations but not on the $\left\langle \Delta_{\kappa} \Delta_{\kappa} \right\rangle$ contributions a pattern of zeroes is generated in the GNC spectrum.
\end{itemize}

The lensing potential for the mode $k = 10 \HH_0$ (\emph{upper left panel} of \autoref{fig:HighCrossBin}) shows that braiding causes an increase of the lensing potential in the Dark Energy dominated era w.r.t. $\Lambda$-GR while the running reduces it. We already addressed this feature and that it directly translates to an increase/decrease of the lensing contributions in the GNC's. Even though the effects of the adopted modified gravity models are only very small for redshifts $z \gtrsim 4.5$, the GNC's on these redshifts are still affected by the modifications: the lensing potential multiplied by a background-depended factor is integrated along the line of sight and therefore also receives contributions from epochs at which the deviations to $\Lambda$-GR are significant.

In the case where the two redshift bins $z_i = 1.5$ and $z_j = 2$ are considered (\emph{upper right panel}), the bins are relatively close together, such that the local terms can not be neglected on ultra-large scales. In $\Lambda$-GR this negative contribution overcomes the positive lensing terms for $\ell < 8$. After that the lensing convergence is the dominant terms up to scales $\ell \sim 100$, where finally the negative contribution from the cross-correlation of the lensing and local terms takes over. Hence, there is another change of sign in the signal on this scale. The latter effect can also be observed for the correlation of higher redshift bins which are well separated (\emph{lower panels}). The first change of sign is not present in these cases as the local terms make barely any contribution in these configurations due to the wide redshift separation of the bins. The oscillations of the signal after the sign crossing are due to the BAOs in the local signal. For the correlation of the bins $z_i = 2.5$ and $z_j = 6$, the scale where the lensing and the local-lensing terms become of the same order and cancel out is shifted to the BAO scales, which induces a richer cancellation pattern in the GNC's.

The signatures of modified gravity models can drastically change the amplitudes of the total signal in high redshift bin configurations that include a change of sign in the signal. As analysed in section \ref{sec:GNC}, the effects integrated along the line of sight, i.e. lensing convergence, Shapiro time-delay and the ISW effect, are modified stronger than local terms (density, RSD and Doppler terms). In models with positive braiding the correlation between the lensing term and the local effects are barely modified while the pure lensing terms get enhanced. Even for correlations between high redshift bins, e.g., $z_i=5$ and $z_j = 8$, this enhancement can become of order 50\% on the largest scales and about 20\% on scales $\ell \sim 100$. Due to this enhancement the scale on which the lensing-local contribution overcomes lensing convergence is shifted towards higher $\ell$ (\emph{upper right} and \emph{lower left panel} of \autoref{fig:HighCrossBin}). It is also interesting that the sign change in $\Lambda$-GR on ultra-large scales for $z_i = 1.5, \, z_j =2$ (\emph{upper right}) is not present for the braiding model. This is due to the fact that the enhancement of the lensing convergence causes its absolute value to overcome the one from the local terms and therefore the signal is dominated by the lensing contribution already on ultra-large scales up to the point where the lensing-local correlation takes over.

For models with a running of the effective Planck mass we can also see modifications in the high redshift bin correlations. A positive run rate of the effective Planck mass reduces the lensing potential and therefore the lensing contribution in the GNC's (\emph{upper left panel} of \autoref{fig:HighCrossBin}). It causes the lensing to overcome the local contribution for larger $\ell$ (in the case of $z_i = 1.5, \, z_j = 2$) and the lensing term is cancelled out by the local-lensing contribution for smaller $\ell$ w.r.t. $\Lambda$-GR (\emph{upper right} and \emph{lower left panel}).

\section{GNC-CMB temperature correlation in Horndeski gravity}
\label{sec:CMB-LSS}

We have found that the relativistic effects involving integrals along the line of sight deviate significantly from the $\Lambda$-GR across the models under study. Fortunately, it is possible to isolate one of such effects by considering the cross-correlation of LSS with the CMB temperature anisotropies. This measurement is dominated by the ISW effect in the CMB, which has the same dependence on the potentials as the ISW effect in the GNC's. This provides a perfect set-up to test the viability of modified gravity models: in fact, our analysis in \autoref{sec:GNC} has shown that the ISW effect is the effect which is most sensitive to the underlying theory of gravity. We start by giving a short overview of the theoretical foundations. Then we explore which regions of the parameter space of the Bellini-Sawicki parameterization predict viable values for the CMB-LSS correlation. The measurements we use have been obtained by Ferraro, Sherwin and Spergel \cite{Ferraro:2014msa} using data from the WISE survey \cite{Wright:2010qw}.

The time variation of the Bardeen potentials causes a fractional variation in the CMB temperature \cite{Sachs:1967er} which is given by
\begin{equation}
\left( \frac{\Delta T}{T} \right)_{ISW} = \int d\tau \, (\Phi'+\Psi') \, .
\end{equation}
The potentials are constant in the matter dominated era and therefore there are no contributions to the ISW effect from this time. It can be only caused by post-recombination radiation, where the matter over-density was very small, or in the Dark Energy dominated era ($z \lesssim 1$). Therefore the major impact on the ISW effect comes from ultra-large scales.
Even though it is sub-dominant in the CMB temperature-temperature anisotropy spectrum, the ISW effect can be isolated considering the cross-correlation between the CMB temperature and galaxy number counts. The angular power spectrum of this cross-correlation can be computed by assuming that the ISW effect is the only contribution at low redshifts, such that:

\begin{equation}
C_{\ell}^{T g} \approx 4 \pi \int \frac{dk}{k} \mathcal{P}_{\mathcal{R}}(k) \Delta_{\ell}^{\mathrm{ISW}} \Delta_{\ell}^{\mathrm{Den}_i} \, ,
\end{equation}
where
\begin{equation}
  \Delta_{\ell}^{\mathrm{ISW}} \equiv \int_0^{\tau_0} d\tau\; e^{-\eta(\tau)} S_{(\Phi'+\Psi')}(k,\tau)\; j_{\ell}\left(k(\tau_0-\tau)\right) \;,
\end{equation}
and $S_{\Phi'+\Psi'}(k,\tau)$ is the source function of the ISW effect, defined in \autoref{sec:transfer}.
We introduced the optical depth $\eta$, and the density transfer function $\Delta_{\ell}^{\mathrm{Den}_i}$, also defined in \autoref{sec:transfer}.
For simplicity and since we are only interested in qualitative changes rather than in a consistent measurement, we neglect galaxy bias fixing $b(z)=1$. The galaxy redshift distribution $dN/dz$ is normalized such that $\int dz' \frac{dN}{dz'} = 1$.

The viability of a gravity model can be tested through CMB-LSS correlations by constraining the \emph{CMB-LSS relative amplitude}
\begin{equation}\label{eq:isw_amplitude}
 \mathcal A = \frac{\sum_{\ell} C^{ T g}_l (MG)}{\sum_{\ell} C^{ T g }_l(\text{ref})}
\end{equation}
defined with respect to the Planck best fit model \cite{Ade:2015xua}. Analysing WISE data \cite{Wright:2010qw} Ferraro et al. \cite{Ferraro:2014msa} found it to be $\mathcal{A} = 1.24 \pm 0.47$.
This measurement can be used to further rule out Covariant Galileon models, as they would predict a negative correlation \cite{Barreira:2014jha}.
To see if the parameterized Horndeski models we used can also be constrained or ruled out by this observation we added the CMB-LSS correlations to {\tt hi\_class}.
In the following we will compute the amplitude $\mathcal{A}$ using a tophat window function centered around $z=0.3$ (the central redshift of WISE galaxies) with bin width $\Delta = 0.05$.  As we are interested in a rough estimate, given the relatively small width of the redshift bin we neglect the evolution of galaxy bias and of the galaxy selection function, and assume that they factor out in the ratio.
Also, since \autoref{eq:isw_amplitude} was constrained in \cite{Ferraro:2014msa} by considering all the WISE galaxies up to $z=1$, caution should be taken when comparing our results to their measurement.

\begin{figure}
 \centering
  \begin{minipage}{0.49\linewidth}
    \centering
    \subfloat [Galileons]
   {\includegraphics[width=\textwidth]
   {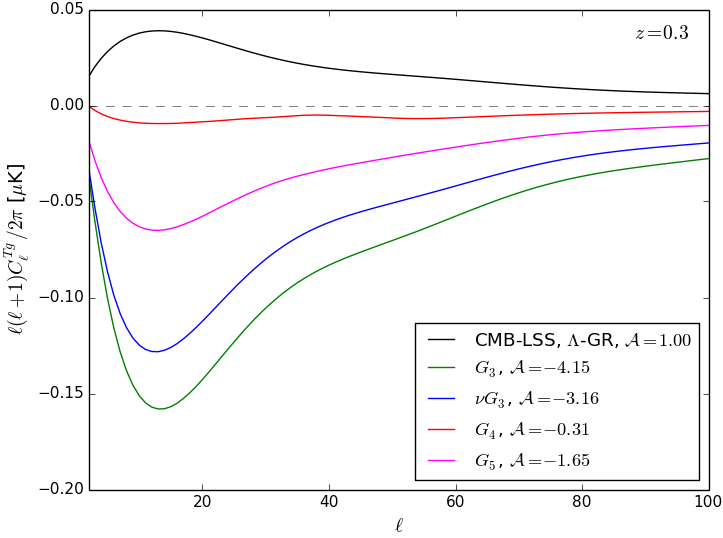}} \\
  \end{minipage}
  \hfill
 \begin{minipage}{0.49\linewidth}
    \centering
 	\subfloat[parameterized Horndeski]
	{\includegraphics[width=\textwidth]
	{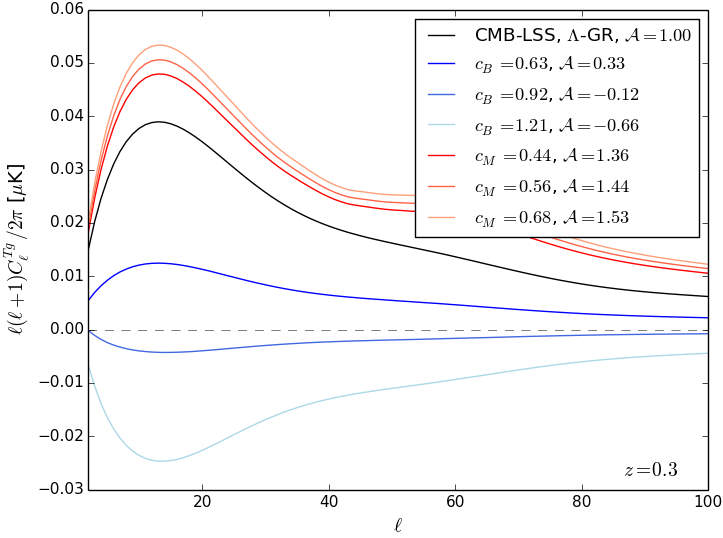}} \\
  \end{minipage}

 \caption{Cross-correlation between large scale structure and CMB temperature anisotropies for Galileon models (\emph{left}) and parameterized Horndeski models (\emph{right}) in comparison to $\Lambda$-GR for redshift $z=0.3$ with tophat window function and bin width $0.05$; note that $c_K = 1$ in all parameterized models.
  \label{fig:CMB-LSS} }
\end{figure}

\autoref{fig:CMB-LSS} shows the CMB-LSS cross-correlation for Covariant Galileons and parameterized models for redshift $z=0.3$ and their corresponding values for the amplitude $\mathcal{A}$. This plots are in perfect agreement with our previous results. The CMB-LSS cross-correlation is enhanced for models with a positive run rate of the effective Planck mass, i.e. $c_M > 0$, as the ISW effect is increased with increasing $c_M$. In models with positive braiding the signal is reduced w.r.t. $\Lambda$-GR as the ISW effect is decreased by increasing $c_B$.
We have omitted models with only kineticity or tensor speed excess from the plots since they barely influence the signal, as expected from the ISW effect analysis in the GNC's.

In $\Lambda$-GR, during the DE dominated epoch Bardeen potentials decay in time.\footnote{Note that since for $\Lambda$-GR we have $\Psi+\Phi<0$, this means $\Psi'+\Phi'>0$.} Therefore, the CMB temperature fluctuation (dominated by the ISW term) is positively correlated to the galaxy over-density. The time-variation of the Bardeen potentials is significant only for CMB photons that travelled along potential wells (or hills) sufficiently extended in redshift, leading to a more important CMB-galaxy correlation on scales comparable to the Hubble horizon and decreasing gradually towards smaller scales.
This is confirmed by the $\Lambda$-GR case shown in \autoref{fig:CMB-LSS}.
The dynamics of Galileons and parameterized Horndeski models is more complex and leads in general to non-trivial scale-dependent deviations from $\Lambda$-GR.

\begin{figure}[t!]
 \centering
  \begin{minipage}{0.49\linewidth}
    \centering
     \subfloat [$c_K=1,\quad c_T=0$]
   {\includegraphics[width=\textwidth]
   {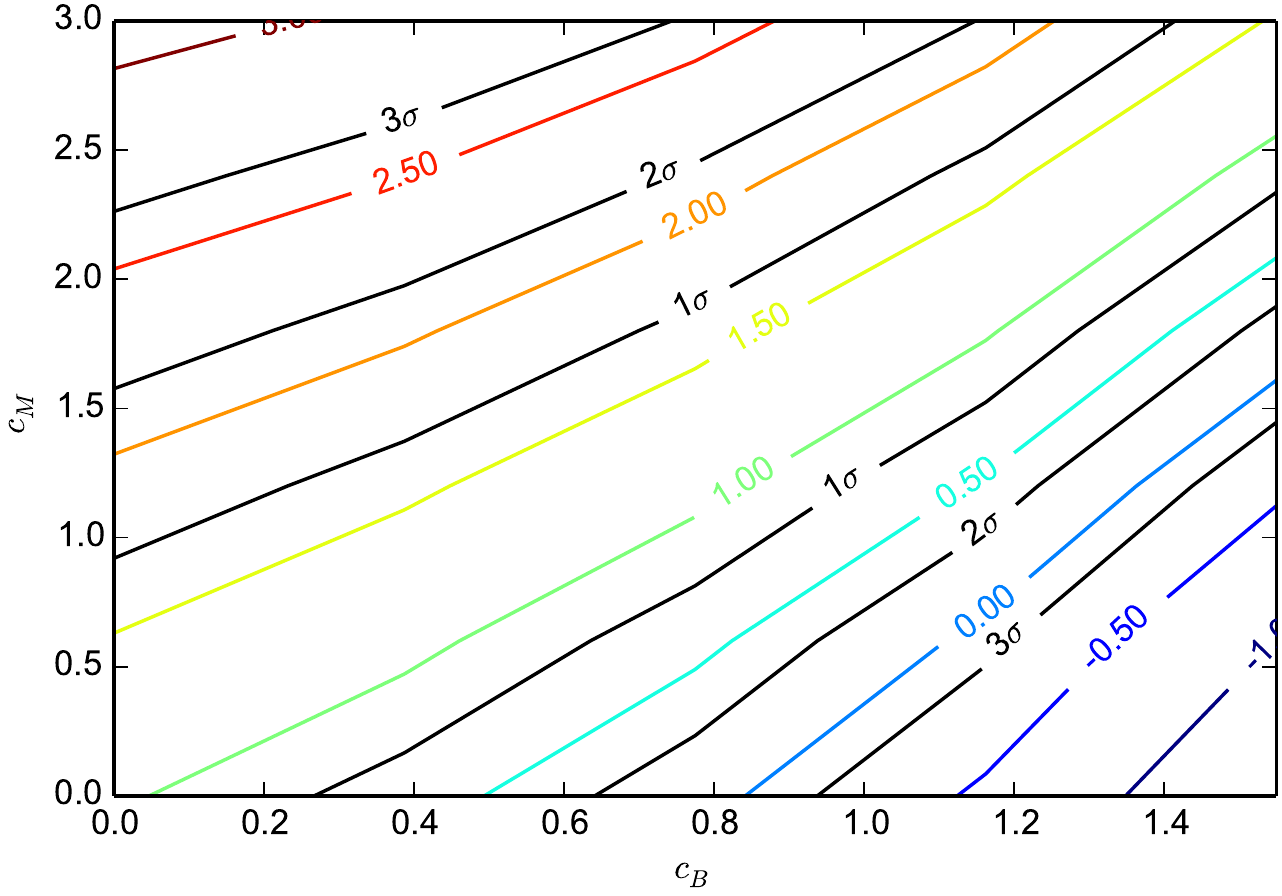}}
  \end{minipage}
  \begin{minipage}{0.49\linewidth}
    \centering
   \subfloat [$c_K=1,\quad c_B=0$]
   {\includegraphics[width=\textwidth]
   {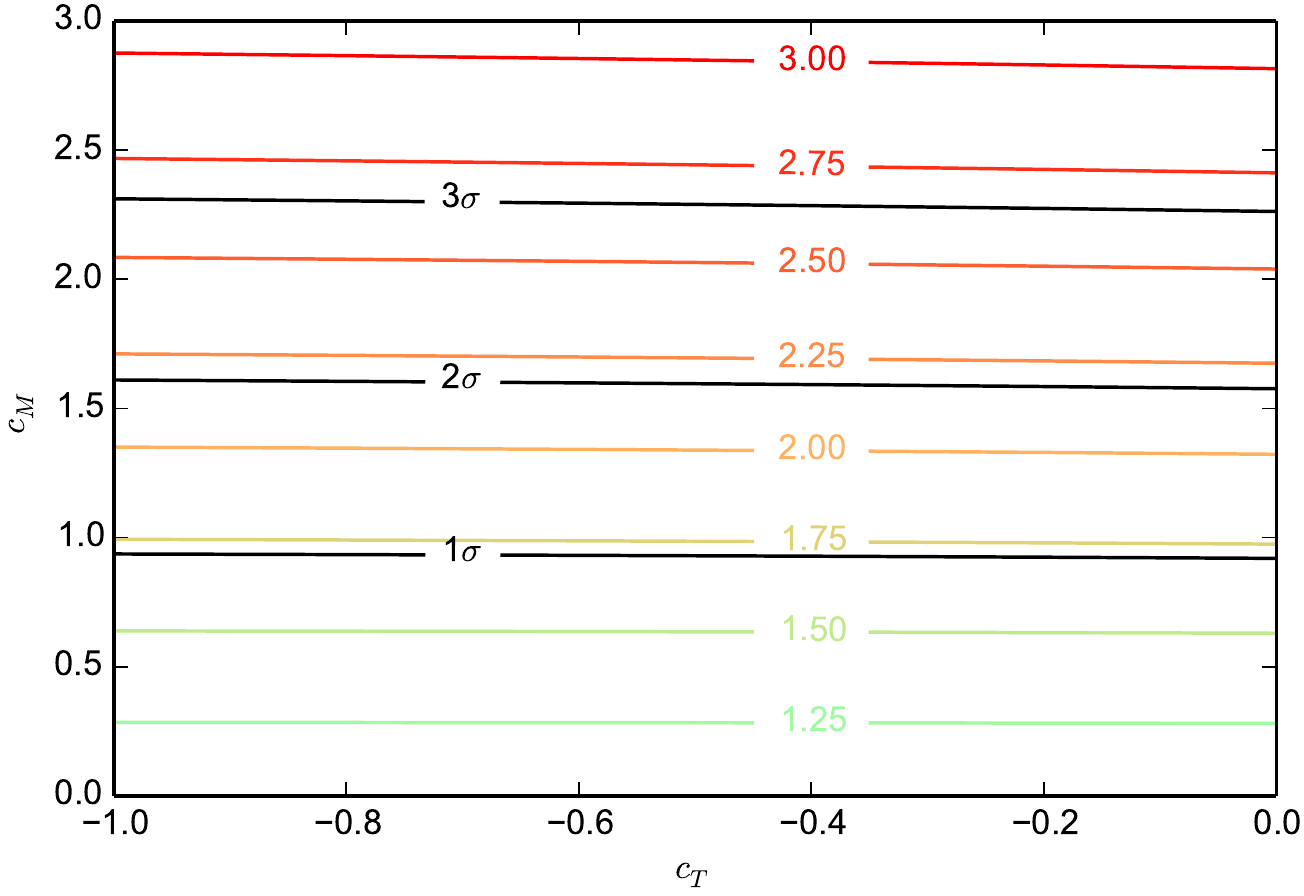}}
  \end{minipage}

 \caption{Cross-correlation between large scale structure and CMB temperature anisotropies for the Bellini-Sawicki parameterization (cf. section \ref{sec:param_bell_sawi}) around the Planck best fit values for $\Lambda$-GR. Colored lines show the values of the relative CMB-LSS relative amplitude (\ref{eq:isw_amplitude}); black lines show the 1,2 and 3$\sigma$ deviations with respect to the measured value $\mathcal{A}=1.24\pm 0.47$ \cite{Ferraro:2014msa}. Note that large values of the braiding can neutralize the ISW-LSS cross-correlation or even turn it into an anti-correlation.
  \label{fig:CMB-LSS_bounds} }
\end{figure}

The \emph{left panel} of \autoref{fig:CMB-LSS_bounds} shows the value of the CMB-LSS amplitude (\autoref{eq:isw_amplitude}) in the $c_B-c_M$ plane of parameterized Horndeski models. The effect of the braiding is to lower the amplitude, eventually leading to negative values of $\mathcal A$ for $c_B\gtrsim 0.8$. The running of the effective Planck mass has the opposite effect, increasing the signal beyond the $\Lambda$-GR prediction and doubling its value for $c_M\gtrsim 1.5$. The two parameters have the opposite effect, leading to net cancellation roughly coinciding with $c_M\approx 2 c_B$. This degeneracy in the ISW-LSS prediction leads to a relatively large region of the parameter space compatible with the measured value of $\mathcal{A}$.
Note that the degeneracy would allow one to distinguish first generation of scalar-tensor theories (i.e. Jordan-Brans-Dicke, $f(R)$, quintessence, K-essence) in which $\alpha_B=-\alpha_M$ from Horndeski models with more general $\alpha_B,\alpha_M$.

The influence of a tensor speed excess gets substantial when combined with braiding or running, even though it has no significant impact on the CMB-LSS correlation if $\alpha_B$ and $\alpha_M$ are both zero.
In combination with the considered braiding model, $c_B = 1.21$, a negative tensor speed excess, e.g. $c_T = -0.98$, causes the CMB-LSS correlation to be reduced at low redshifts ($z \sim 0.3$) compared to a model with $c_B = 1.21$ but $c_T=0$. This reduction at $z = 0.3$ is of order 5\% on the scale $\ell = 50$. For high redshifts ($z \gtrsim 1$) the tensor speed excess causes an enhancement of the signal and the effect is considerably stronger. In the given example this increase is about 95\% at redshift $z=2.5$ and scale $\ell = 50$.

In the case of a combination of $c_T$ and $c_M$ the behaviour is similar. A negative value of $c_T$ just slightly reduces the CMB-LSS correlation for low redshifts ($\sim 0.5\%$) but has a larger impact on higher redshifts, where it increases the signal (about $11 \%$ for $z = 2.5$ at scale $\ell = 50$).
This feature can potentially be used to isolate the impact of the propagation speed of gravitational waves by comparing the CMB-LSS cross-correlation at different redshifts, as the impact of $\alpha_T$ is strongly increased for high redshifts.

\section{Conclusions}
\label{sec:Concl}

We have systematically studied the impact of consistent, alternative theories of gravity on LSS observations involving ultra-large scales. Our theoretical landscape is given by models within the Horndeski Lagrangian, a very general theory that encompasses many viable models for Dark Energy and modified gravity. Among these models, we have focused on Covariant Galileons and the minimal parameterization of the dynamics proposed by Bellini and Sawicki, using the \hiclass\, code\footnote{\url{www.hiclass-code.net}} \cite{Zumalacarregui:2016pph} to obtain predictions for the observable quantities in the linear regime.
Our work is the first to explore this regime of structure formation using fully consistent models in this very general class and is complementary to previous studies that use either parameterizations of the solutions (inspired by quasi-static results), cf. \cite{Lombriser:2013aj,Baker:2015bva}, or focus on very specific models \cite{PhysRevD.75.044004, PhysRevD.73.123504,PhysRevD.91.084046}. We also extend previous analyses of relativistic effects in modified gravity by considering correlations of the galaxy distribution at different redshifts.

As a first observable we have considered power spectra of galaxy number counts.
In a relativistic framework, GNC's depend not only on the intrinsic clustering and RSD terms, but also on other effects non-negligible on the largest scales $k\sim \mathcal H$. At least part of those will be detectable by future experiments \cite{Alonso:2015sfa,Fonseca:2015laa}. These contributions to GNC's depend directly not only on the growth of matter, but also on the gravitational potentials, making them direct probes of modifications of gravity. Moreover, the possibility of choosing different redshift and angular configurations (e.g. correlating galaxies in the same or different redshift bins) allows one to separately measure some combinations of the relativistic effects.

Correlations of galaxies at the same redshift are dominated by the density contribution, followed by Kaiser-RSD, Doppler terms/lensing convergence (depending on the angular scale), local potentials, time-delay and the ISW effects \cite{Bonvin:2011bg}. Modified gravity does not significantly alter this hierarchy, in which each contribution is roughly a factor 5-10 smaller than the preceding one, with the exception of swapping the ISW effect and time delay in some models.
The impact of modified gravity is largest on the least dominant contributions, and the discrepancies with respect to $\Lambda$-GR are usually dominated by the density and Kaiser-RSD. The total deviations can become of order $\sim 40$\% for Covariant Galileons in which the expansion history and the gravitational force are modified. For parameterized models with standard expansion histories they are typically small as the modifications to the leading terms, density and RSD, are minor.

Galaxy correlations at different redshift can be used to isolate the contributions which are most sensitive to modified gravity, i.e. the effects that involve integrals along the line of sight.
The lensing convergence can even dominate the signal and therefore the deviations on the largest scales, specially if broad, non overlapping redshift bins are considered \cite{Montanari:2015rga,Cardona:2016qxn}.
It is important to notice that the aforementioned lensing effect is only due to magnification: it does not require measuring galaxy shapes and it is therefore subject to completely different systematics than the usual weak lensing measurements.
With very precise measurements on ultra-large scales (e.g. using multiple tracers), deviations from $\Lambda$-GR due not only to the lensing contribution, but also to the other sub-dominant relativistic effects might be relevant \cite{Alonso:2015sfa,Fonseca:2015laa,Bonvin:2013ogt}.

In alternative theories of gravity, the balance between the contributions to different redshift correlations is modified with respect to $\Lambda$-GR. The strong enhancement of the integrated terms can lead to several signatures: in parameterized Horndeski models the contribution from the cross-correlation of lensing with the other integrated terms (i.e. lensing-ISW and lensing-time-delay) can overcome the contribution from the Shapiro time-delay and become the third largest for the redshifts we have considered. The two leading terms are the lensing convergence itself and the cross-correlation of lensing with local terms.
In Galileons models the lensing-integrated terms correlation can even overcome the latter contribution to be the second largest effect and by itself become larger than the total signal in standard gravity.
Although the signal is small, effects such as the lensing convergence will be relevant for future surveys such as Euclid \cite{Montanari:2015rga} and could provide a novel test of gravity.
The integrated effects have the additional advantage of being sensitive to a broad range of radial scales.
This dependence can potentially allow LSS observations at high redshifts (for which $\Omega_{de}(z)\ll 1$) to obtain information about Dark Energy, in a manner analogous to proposals for weak lensing in 21cm surveys \cite{Zhang:2005eb,Lu:2009je,Masui:2009cj}.
In addition, higher redshift observations contain more modes comparable to the Hubble horizon in that epoch and are likely to make these effects more important.

Cross-correlations of GNC's at high but different redshift may provide strong constraints on modified gravity, even when that redshifts corresponds to eras in which GR is effectively recovered. This is due to the integrated nature of the dominant effects and can lead to deviations exceeding 50\% on the largest scales.
In configurations that involve a change of sign of the total signal, the difference between $\Lambda$-GR and modified gravity models can be of several orders of magnitude. The sign change is due to the cancellation of the two dominant effects, the lensing convergence (positive) and the correlation between lensing and local terms (negative). In modified gravity the effects on the first contribution are more drastic, and the scale on which both effects cancel out is shifted. By choosing the considered redshift bins appropriately the scale of cancellation can coincide with the BAO scales leading to an oscillation of the total signal around zero on these angular scales.
It is worth to note that, unfortunately, intensity mapping surveys are not affected by lensing convergence at first order. This effect, that also happens to the CMB, is due to the fact that intensity mapping measures surface brightness, which is conserved by lensing \cite{Hall:2012wd}. Profiting from this effect to test gravity in regular galaxy surveys would still require good control of the lensing magnification and galaxy bias, which are degenerate with these effects.

We investigated in detail the effects of non-standard gravity on each contribution to the GNC in the context of the Bellini-Sawicki parameterization \cite{Bellini:2014fua}.
While varying the kineticity parameter and the tensor speed excess alone does not bring substantial changes to the GNC's, changes in the running of the effective Planck mass and of the braiding parameter lead to significant deviations.
A positive run rate causes an effective increase of matter clustering at late times, but a decrease of the lensing potential w.r.t. $\Lambda$-GR.
In pure braiding models both metric potentials are enhanced (but without leading to an anisotropic stress), which is particular interesting for different redshift correlations dominated by integrated terms.
Further non-trivial dependencies appear in the GNC's when the Horndeski parameters are varied jointly.

We have found that the ISW effect is the most sensitive to the theory of gravity, deviating from $\Lambda$-GR by as much as 90\% for parameterizations and up to 2000\% for Covariant Galileons. While this effect is negligible in GNC's, it can be isolated by cross-correlating galaxy catalogues with CMB temperature maps. This correlation has been measured to be positive and in good agreement with the $\Lambda$-GR prediction. However, in the majority of the models we have considered the ISW effect can turn to an anti-correlation between CMB temperature and LSS, as it is the case for all best-fit Galileons and models with significant braiding. This dramatic difference exemplifies the power of combining CMB and LSS to probe the nature of gravity.

We can gain further insights into the connection between the ISW effect and the fundamental properties of gravity by means of parameterized models.
Our results show that an increase in the effective Planck mass with time, $\alpha_M$, tends to increase the magnitude of the correlation, while a kinetic mixing between the metric perturbations and the scalar field, $\alpha_B$, tends to decrease its value and can potentially make it negative. This naturally leads to a degeneracy in the direction $\alpha_B\approx 2 \alpha_M$ on which $C_\ell^{T g}$ remains close to the standard value.
The anomalous propagation of gravitational waves, $\alpha_T$, can be isolated from other modified gravity effects by comparing the CMB-LSS correlation for different redshifts. $\alpha_T$ in combination with $\alpha_M$ causes a slight decrease of the signal on low redshifts and an increase at higher redshifts compared to models with $\alpha_T=0$, while we observe the inverse trend in models with braiding.

Future galaxy surveys will hunt for relativistic effects in cosmological observables as yet another confirmation of Einstein's theory. These observations offer an excellent opportunity to test gravity in a regime that is naturally at the same dynamical scale as cosmic acceleration. We have studied the impact of consistent modifications of gravity in this regime, paving the way to more specific studies of how these effects can be observationally detected. A very promising technique in this direction is the use of multiple tracers of the LSS distribution, either to reduce cosmic variance or to measure the anisotropic correlation function, directly related to sub-dominant contributions.
The cross-correlation between LSS and CMB temperature allows one to isolate the ISW effect and it provides a clean, direct test of the properties of gravity. Other cross-correlations between different field are also possible and might provide complementary signatures \cite{Pullen:2015vtb}.

Another avenue for further development can be pursued by considering more general theoretical frameworks. Beyond Horndeski theories of the $G^3$ type \cite{Gleyzes:2014dya} have characteristic signatures that vanish in the quasi-static regime \cite{Lombriser:2015cla}. This might be more critical to test degravitation models that address the cosmological constant problem, often leading to signatures only on curvature regime comparable to the residual $\Lambda$ \cite{Kaloper:2015jra}.
These observational prospects and theoretical developments anticipate a new era in which large-volume galaxy surveys will bring in new insights into fundamental physics and the nature of gravity.

\paragraph{Acknowledgements:}

We are very grateful to Luca Amendola for initial discussions and many useful suggestions along the way, David Alonso and Ruth Durrer for comments on the draft, Daniele Bertacca, Enea Di Dio, Uros Seljak and Alvise Raccanelli for discussions of relativistic effects, as well as Simone Ferraro and Laura Taddei for for discussions on the ISW effect.
JR and MZ are partially supported is supported by DFG through the grant TRR33 ``The Dark Universe''.  JR acknowledges support from Stockholm University and the Oskar Klein Center.

\appendix

\section{Brans-Dicke theory}\label{sec:BD_theory}
The Brans-Dicke \cite{Brans:1961sx} theory is specified by the following choice of the Horndeski functions
\begin{equation}
 G_2 = M_p^2\frac{\omega}{\phi}X -\Lambda\,,\quad G_4 =  \phi\frac{M_p^2}{2}\,,\quad G_3=G_5=0\,,
\end{equation}
where the scalar field $\phi$ is dimensionless and $M_p = (8\pi G)^{-1/2}$ is the reduced Planck mass.
Since the Brans-Dicke theory has no screening mechanism, we set the initial conditions on the scalar field so that $\phi=1$ at $z=0$.
$\Lambda$-GR is recovered for $w \rightarrow \infty$. Using CMB data from Planck the Brans-Dicke parameter was constrained to be $\omega>692$ at 99\% confidence level \cite{Avilez:2013dxa}. Further constraints on the Brans-Dicke parameter using Planck or WMAP data can be found in \cite{Li:2013nwa, Wu:2009zb, Acquaviva:2004ti}.

\begin{figure}[t!]
 \centering
  \begin{minipage}{0.49\linewidth}
    \centering
    \subfloat [auto \& cross-correlations]
	{\includegraphics[width=\textwidth]
	{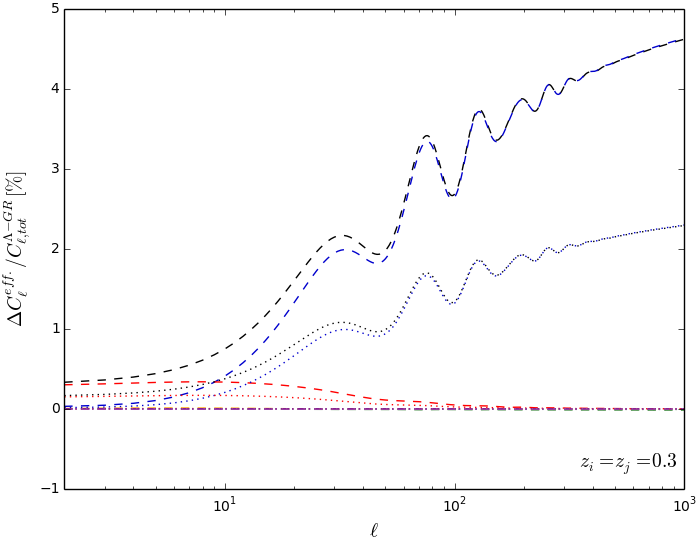}}
  \end{minipage}
  \hfill
 \begin{minipage}{0.49\linewidth}
    \centering
 	\subfloat[auto correlations only]
   {\includegraphics[width=\textwidth]
   {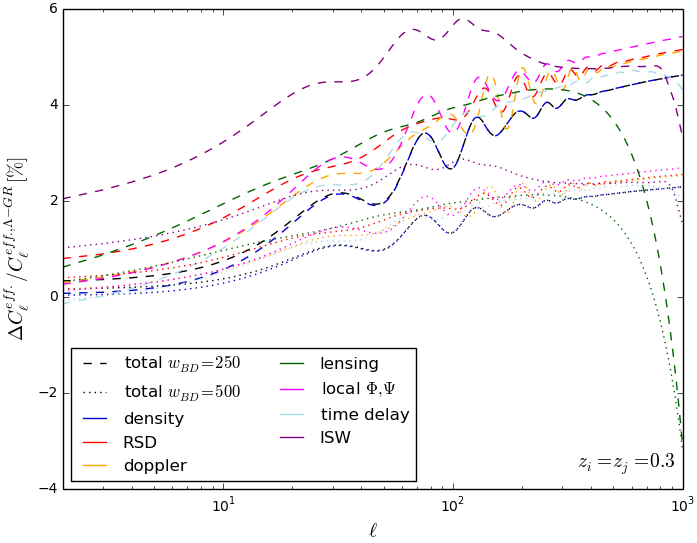}}
  \end{minipage}
 \caption{Fractional deviations of total galaxy number count power spectrum and relativistic effects for two Brans-Dicke models from $\Lambda$-GR for $z_i=z_j = 0.3$ and tophat window function with bin width $0.05$. \emph{Left}: Fractional deviations of relativistic effects including their cross-correlation with density w.r.t. the total signal in $\Lambda$-GR. \emph{Right}: The fractional deviation of each effect w.r.t. the corresponding effect in $\Lambda$-GR.
  \label{fig:BDAutoBin}   }
\end{figure}

We investigate the matter and metric perturbations for two Brans-Dicke models with the parameters $w=500$ and $w = 250$.
Driving the Brans-Dicke parameter further away form $\Lambda$-GR (i.e. decreasing $w$) the matter power spectrum gets enhanced for sub-horizon modes. The same applies to the Bardeen potentials and therefore the lensing potential.
In analogy to the previous results this translates to an increase of the local terms in the GNC's (due to the enhancement of matter perturbations) and of the effects directly depending on the metric perturbations. This can be seen in \autoref{fig:BDAutoBin} where we show the fractional deviations of the GNC's for redshift $z=0.3$ from $\Lambda$-GR for two Brans-Dicke models. Neither the total deviations nor the deviations of the single contributions exceed 5\% if models consistent with the constraint $\omega>692$ are considered. The same applies to the correlation of different redshift bins (apart from the scales on which the signal crosses zero) and to the correlation of the GNC's with CMB temperature anisotropies.

\section{Transfer functions}
\label{sec:transfer}
We compute the linear transfer functions of relativistic number counts with \hiclass\footnote{\url{http://hiclass-code.net/}} \cite{Zumalacarregui:2016pph} a modified version of the {\sc CLASS} code\footnote{\url{http://class-code.net/}} \cite{Blas:2011rf}, based on the implementation introduced in {\sc CLASSgal}\footnote{\url{http://cosmology.unige.ch/content/classgal}} \cite{DiDio:2013bqa}.
For more details see, e.g., \cite{Bonvin:2011bg,Challinor:2011bk,DiDio:2013bqa}.

The transfer functions read:
\begin{eqnarray}
\Delta_{\ell}^{\mathrm{Den}_i} &=& \int_0^{\tau_0} d\tau W_i \, b(z) S_\mathrm{\delta} \, j_{\ell} \nonumber \\
\Delta_{\ell}^{\mathrm{Len}_i} &=& \ell(\ell+1) \int_0^{\tau_0} d\tau \, W^\mathrm{L}_i \,  S_{\Phi+\Psi} \, j_{\ell} \nonumber \\
\Delta_{\ell}^{\mathrm{V}1_i} &=& \int_0^{\tau_0} d\tau \, W_i \left[ 1 \! + \! \frac{H'}{aH^2} \! + \frac{2 -5s }{(\tau_0-\tau)aH }  \!  +5s - f_{\rm evo}  \right] \frac{S_{\Theta}}{k} \, \frac{d j_{\ell}}{dx}  \nonumber \\
\Delta_{\ell}^{\mathrm{V}2_i} &=& \int_0^{\tau_0} d\tau \, W_i \left(f_{\rm evo} -3 \right)aH \frac{S_{\Theta}}{k^2} \, j_{\ell} \nonumber \\
\Delta_{\ell}^{\mathrm{V}3_i} &=& \int_0^{\tau_0} d\tau \, W_i \left( \frac{1}{aH} \right) S_{\Theta} \, \frac{d^2j_{\ell}}{dx^2} \nonumber \\
\Delta_{\ell}^{\mathrm{G}1_i} &=& \int_0^{\tau_0} d\tau \, W_i \left[c_{\rm lpot}+(c_{\rm lpot}-c_{\rm isw})\left( 1+\frac{H'}{aH^2} +  \frac{2 -5s}{(\tau_0-\tau) aH} + 5s-f_{\rm evo}  \right)\right] S_\Psi \, j_{\ell} \nonumber \\
\Delta_{\ell}^{\mathrm{G}2_i} &=& \int_0^{\tau_0} d\tau \, W_i  \left[ (c_{\rm lpot}-c_{\rm isw})(-2+5s)  - c_{\rm isw}\left(3+\frac{H'}{aH^2} + \frac{2 -5s }{(\tau_0-\tau)aH } -f_{\rm evo}  \right)  \right] S_\Phi \, j_{\ell} \nonumber \\
\Delta_{\ell}^{\mathrm{G}3_i} &=& c_{\rm lpot} \int_0^{\tau_0} d\tau \, W_i \, \left( \frac{1}{aH} \right) S_{\Phi'} \, j_{\ell} \nonumber \\
\Delta_{\ell}^{\mathrm{G}4_i} &=& c_{\rm std} \int_0^{\tau_0} d\tau \, W_i^{\mathrm{G}4} \,  S_{\Phi+\Psi} \, j_{\ell} \nonumber \\
\Delta_{\ell}^{\mathrm{G}5_i} &=& c_{\rm isw}\int_0^{\tau_0} d\tau \, W_i^{\mathrm{G}5} \, S_{(\Phi+\Psi)} k\, \frac{dj_{\ell}}{dx} ~,
\label{eq:delta_terms}
\end{eqnarray}
where for the integrated terms we defined
\begin{eqnarray}
W_i^\mathrm{L}(\tau) &=& \int_0^\tau \!\! d\tilde{\tau} W_i(\tilde{\tau}) \left( \frac{2-5s}{2} \right) \frac{\tau-\tilde\tau}{(\tau_0-\tau)(\tau_0-\tilde\tau)} \nonumber \\
W_i^{\mathrm{G}4}(\tau) &=&  \int_0^\tau \!\! d\tilde{\tau} W_i(\tilde{\tau})  k\,  \frac{2 -5s}{\tau_0-\tilde\tau} \\
W_i^{\mathrm{G}5}(\tau) &=& \int_0^\tau \!\! d\tilde{\tau} W_i(\tilde{\tau}) \left[1+\frac{H'}{aH^2} +  \frac{2 -5s}{(\tau_0-\tilde\tau) aH} + 5s-f_{\rm evo} \right]_{\tilde{\tau}}~. \nonumber
\end{eqnarray}
Here the source functions $S_X(k,\tau)$ are linear combinations of transfer functions $A(\tau,k)\equiv A(\tau,\bk)/\mathcal{R}(\tau_{\rm ini},\bk)$, where $A(\tau,\bk)$ is any perturbation with adiabatic initial conditions, and $\mathcal{R}(\tau_{\rm ini},\bk)$ is the curvature perturbation at some initial time $\tau_{\rm ini}$ such that the corresponding mode is super horizon $k\tau_{\rm ini}\ll1$.
We integrate over a window function $W(z_i)$ centered at redshift $z_i$ that encodes information about the redshift resolution (typically a tophat for spectroscopic resolution, or a Gaussian with variance given by photometric resolution) and about the distribution of galaxies per redshift and per solid angle $dN/dz/d\Omega$.
We omitted the arguments $k$ for the transfer functions $\Delta$, $(\tau,k)$ for the source functions $S_X$, $x\equiv k(\tau_0-\tau)$ for the Bessel functions $j_{\ell}$ ($\tau_0$ being the conformal time today), and $\tau$ for selection and background functions.
We also introduced the velocity source function $S_\Theta(\tau,k)$ given by $\Theta(k) \equiv kV(k)$, where $V(k)$ is the velocity perturbation in the Newtonian gauge.
Primes indicate derivatives with respect to conformal time.
In the following paragraph we describe the newly introduced coefficients $c_{\rm isw}$, $c_{\rm std}$ and $c_{\rm lpot}$ added to identify the ISW, Shapiro time-delay and other local terms depending on the Bardeen potentials, respectively.
These coefficients allow us a more physical description of the relativistic terms $\Delta_{\ell}^{\mathrm{G}i}$, while still splitting the contributions in the code according to their source function and number of derivatives of spherical Bessel functions.

The different contributions shown above correspond to the density in comoving gauge $\delta_{\rm co}$ (``Den''), lensing convergence $\kappa$ (``Len''), Doppler (``V1''-``V2''), redshift-space distortions in the Kaiser approximation (``V3'') and terms depending on gravitational potential (``G1''-``G5''), respectively.
The separation of different contributions is consistent with the {\sc CLASS} code.
In particular, it differs from the previous {\sc CLASSgal} implementation as, for numerical convenience, the time derivative $\Psi'$ (requiring numerical derivatives of perturbation equations) has been integrated by parts.
The corresponding expressions for the relativistic transfer functions (other than lensing convergence) presented in {\sc CLASSgal} \cite{DiDio:2013bqa} are
\begin{eqnarray}
\Delta_{\ell}^{\widetilde{\mathrm{G}1}_i} &=& c_{\rm lpot} \int_0^{\tau_0} d\tau \, W_i \left( 2+\frac{H'}{aH^2} +  \frac{2 -5s}{(\tau_0-\tilde\tau) aH} + 5s-f_{\rm evo}  \right) S_\Psi \, j_{\ell} \nonumber \\
\Delta_{\ell}^{\widetilde{\mathrm{G}2}_i} &=& c_{\rm lpot} \int_0^{\tau_0} d\tau \, W_i (-2+5s) S_\Phi \, j_{\ell} \nonumber \\
\Delta_{\ell}^{\mathrm{G}3_i} &=& c_{\rm lpot} \int_0^{\tau_0} d\tau \, W_i \, \left( \frac{1}{aH} \right) S_{\Phi'} \, j_{\ell} \nonumber \\
\Delta_{\ell}^{\mathrm{G}4_i} &=& c_{\rm std} \int_0^{\tau_0} d\tau \, W_i^{\mathrm{G}4} \,  S_{\Phi+\Psi} \, j_{\ell} \nonumber \\
\Delta_{\ell}^{\widetilde{\mathrm{G}5}_i} &=& c_{\rm isw}\int_0^{\tau_0} d\tau \, W_i^{\mathrm{G}5} \, S_{(\Phi'+\Psi')} \, j_{\ell} ~,
\label{eq:delta_terms_classgal}
\end{eqnarray}
where a tilde $\widetilde{Gi}$ indicates expressions proper of {\sc CLASSgal}.
Equations (\ref{eq:delta_terms}), implemented in {\sc CLASS}, are recovered once we integrate by parts $\Delta_{\ell}^{\widetilde{\mathrm{G}5}_i}$ (neglecting boundary terms since they vanish as $\tau\to0$ and are unobservable for $\tau=\tau_0$) and redefining consistently $\Delta_{\ell}^{\widetilde{\mathrm{G}1}_i}$ and $\Delta_{\ell}^{\widetilde{\mathrm{G}2}_i}$ such that the sum of these three terms is unchanged.
It is worth noting that the time derivative $\Phi'$ entering in $\Delta_{\ell}^{\mathrm{G}3_i}$ can usually be obtained analytically from perturbation equations, so it does not represent a numerical issue.
Furthermore, when integrating by parts, to keep track of the terms representing local potential contributions, Shapiro time-delay and ISW, we introduced (inspired by \cite{Lesgourgues:2013bra}) the coefficients $c_{\rm lpot}$, $c_{\rm std}$, and $c_{\rm isw}$, respectively.

\bibliographystyle{JHEP}
\bibliography{relativistic}

\end{document}